\documentclass[aps,prc,letterpaper,11pt,twoside,tightenlines,nofootinbib,showpacs,preprint]{revtex4-1}
\usepackage{graphicx}
\usepackage[sort&compress]{natbib}
\usepackage{subfigure}
\usepackage{amsmath}
\usepackage{amsfonts}
\usepackage{cancel}
\usepackage{amssymb}
\usepackage{hyperref}
\usepackage{color}

\begin{document}
\arraycolsep1.5pt
\newcommand{\Ima}{\textrm{Im}}
\newcommand{\Rea}{\textrm{Re}}
\newcommand{\mev}{\textrm{ MeV}}
\newcommand{\be}{\begin{equation}}
\newcommand{\ee}{\end{equation}}
\newcommand{\ba}{\begin{eqnarray}}
\newcommand{\ea}{\end{eqnarray}}
\newcommand{\gev}{\textrm{ GeV}}
\newcommand{\nn}{{\nonumber}}
\newcommand{\dtres}{d^{\hspace{0.1mm} 3}\hspace{-0.5mm}}
\newcommand{\rts}{ \sqrt s}
\newcommand{\non}{\nonumber \\[2mm]}

\title{Prediction of a $Z_c(4000)$ $D^* \bar D^*$ state and relationship to the claimed $Z_c(4025)$.}

\author{F. Aceti$^{1}$,  M. Bayar$^{1,2}$, J. M. Dias$^{1,3}$ and E. Oset$^{1}$}
\affiliation{$^{1}$Departamento de F\'{\i}sica Te\'orica, Universidad de Valencia and IFIC, Centro Mixto Universidad de
Valencia-CSIC, Institutos de Investigaci\'on de Paterna, Aptdo. 22085, 46071 Valencia,
Spain\\ \\$^{2}$Department of Physics, Kocaeli University, 41380 Izmit, Turkey\\ \\
$^{3}$Instituto de F\'isica, Universidade de S\~ao Paulo, C. P. 66318, 05389-970, S\~ao Paulo, SP, Brasil.
 }

\date{\today}

\begin{abstract}
After discussing the OZI suppression of one light meson exchange in the interaction of $D^* \bar D^*$ with isospin I=1, we study the contribution of two pion exchange to the interaction and the exchange of a heavy vectors, $J/\psi$ for diagonal transitions $D^* \bar D^*$ and $D^*$ for transitions of $D^* \bar D^*$ to $J/\psi\,\rho$. We find these latter mechanisms weak, but enough to barely bind the system in J=2 with a mass around 4000 MeV, while the effect of the two pion exchange is a net attraction but weaker than that from heavy vector exchange. We discuss this state and try to relate it to the $Z_c(4025)$ state, above the $D^* \bar D^*$ threshold, claimed in an experiment at BES from an enhancement of the $D^* \bar D^*$ distribution close to threshold. Together with the results from a recent reanalysis of the BES experiment showing that it is compatible with a J=2 state below threshold around 3990 MeV, we conclude that the BES experiment could be showing the existence of the state that we find in our approach. 
\end{abstract}

\maketitle

\section{Introduction}
\label{Intro}

The charmonium spectrum of $c \bar c $ states has been enriched with a plethora of new states called $X$,$Y$,$Z$ states, which do not fit in the expected spectrum of ordinary $c \bar c$ quark states \cite{Ali:2011vy,Gersabeck:2012rp,Olsen:2012zz,Li:2012pd}. The theory has followed trend with also a rich offer of possible interpretations, like tetraquarks, or molecular states, and more exotic states \cite{Brambilla:2010cs}. One of the last surprises has been the finding of $Z$ states with isospin $I=1$. In the hidden charm sector a state around $4020$ MeV, called $Z_c(4020)$ and a width of about $8$ MeV has been reported in \cite{Ablikim:2013wzq}, in the $e^+ e^- \to \pi^+ \pi^- h_c$ reaction, looking at the invariant mass of $\pi^{\pm} h_c$. Another BES experiment has found a peak in the $(D^* \bar D^*)^{\pm}$ spectrum close to threshold, which was interpreted in terms of a new resonance with mass around $4025$ MeV and width about $25$ MeV \cite{besexp}. It is unclear whether these two states can be the same, and the quantum numbers are in any case not well determined. The peak seen in the $(D^* \bar D^*)^{\pm}$ spectrum is appealing since in \cite{raquel2} the study of the $D^* \bar D^*$ interaction gave rise to a state with $I=1$ in spin $J=2$. The state appeared around $3920$ MeV, with uncertainties. Actually, we will claim here that it should be much less bound, but that most probably it is related to the peak seen in the
$(D^* \bar D^*)^{\pm}$ in \cite{besexp}. The threshold for $(D^* \bar D^*)^{\pm}$ is $4017$ MeV, so a bound state of $D^* \bar D^*$ should have a smaller energy, while the energy of the state is claimed at $4025$ MeV in \cite{besexp}. Yet, the interpretation of peaks around threshold is always problematic and a source of confusion. Indeed, most often an enhancement of the invariant mass at threshold is an indication of a bound state or resonance below threshold.
There are multiple examples of it. In a similar
reaction, $e^+e^− \to J/\psi D \bar D $ \cite{Abe:2007sya}, a bump close to the threshold
in the $D \bar D$ invariant mass distribution was reported by the Belle collaboration, which was tentatively interpreted as a new resonance.
This peak was, however, interpreted in Ref. \cite{danibump} in terms of a bound $D \bar D $ molecular
state, called $X(3700)$, which had been predicted in Ref. \cite{daniddbar} and later on
has also been reported in other works
\cite{HidalgoDuque:2012pq,Nieves:2012tt,Guo:2013sya,Liu:2010xh,Zhang:2006ix,zzy1}. In a similar way, in Ref. \cite{MartinezTorres:2012du}, a peak seen in the $\phi \omega$
threshold in the $J/\psi \to \gamma \phi \omega$ reaction \cite{Ablikim:2006dw} was better interpreted as a manifestation of the
$f_0(1710)$ resonance, below the $\phi \omega$ threshold, which couples strongly to $\phi \omega$ \cite{gengvec}. More recently
a bump close to threshold in the $K^{∗0} \bar K^{∗0}$
invariant mass distribution, seen in the $J/\psi \to \eta K^{∗0} \bar K^{∗0}$ decay in Ref. \cite{Ablikim:2009ac}, is interpreted in \cite{miguel}
as a signal of the formation of an $h_1$ resonance, predicted in Ref. \cite{gengvec}, which couples mostly to the $K^{*} \bar{ K}^{*}$.

In the same direction as in the previous works, in \cite{alberdd} the experiment of \cite{besexp} was reanalyzed and the enhancement in the $D^* \bar D^*$ invariant mass distribution was found compatible with a state with $J=2$, mass around $3990$ MeV and width around $160$ MeV, although fits with other solutions were also found acceptable. Yet, resonances with mass bigger than the $D^* \bar D^*$ mass were discouraged based on the difficulty to have single channel resonances with energy above threshold. Indeed, it was shown in \cite{junkojuan} that an energy independent potential, smooth in momentum space, could not generate a resonance above the mass of the interacting particles. In this sense, any energy below threshold is preferred, and the $J=2$ solution with mass around $3990$ was proposed as a good candidate to explain the experimental peak. Another reason in favour of this interpretation was that if the state were a $J^P=1^+$ produced in $S$-wave, as assumed in the experimental work \cite{besexp}, it can easily decay into $\pi J/\psi$. This decay channel is the same of the $Z_c(3900)$ \cite{zc3900}. However, while a peak is clearly visible in the
$\pi J/\psi$ invariant mass distribution for the $Z_c(3900)$, no peak is seen around $4025$ MeV (see Fig.~4 of Ref.~\cite{zc3900}).

In the present work we go back to \cite{raquel2} and perform some corrections to update the results of the local hidden gauge to the results of the heavy quark spin symmetry \cite{wise}. On the other hand we also show how for $I=1$, the exchange of light $q \bar q$ states is OZI forbidden, which makes the exchange of light vectors and pseudoscalars cancel when equal masses are taken for them separately, and, because of that, gives a small contribution when real masses are used. In view of this we explore the exchange of two pions, both with and without interaction. The exchange of vector mesons is reduced to the exchange of $J/\psi$ in the $D^* \bar D^*$ diagonal terms, or $D^*$ in the $D^* \bar D^* \rightarrow J/\psi\,\rho$ transition, which makes the potential small, in spite of which we still find it bigger than that of the two pion exchange. Altogether we find a state of the $D^* \bar D^*$ in $I=1$, $J=2$, close to threshold, which, together with the findings of \cite{alberdd}, offers a natural interpretation for the peak observed in \cite{besexp}.
This molecular interpretation would also be supported by QCD sum rules calculations, \cite{Chen:2013omd}, \cite{Khemchandani:2013iwa}, \cite{Cui:2013vfa},\cite{Cui:2013yva}, although the uncertainties between $\pm 105$ MeV and $\pm 280$ MeV in the binding of these works offers only a weak support to our more precise determination of the mass. At the same time, QCD sum rules disfavor the interpretation of the $Z_c(4025)$ as a possible diquark-antidiquark type vector tetraquark states \cite{Wang:2013exa}. A molecular interpretation for this state is also assumed in \cite{valery}, where the coupling to the $D^*\bar{D}^*$ components is evaluated by means of the Weinberg compositeness condition \cite{weinberg} and this picture is used to evaluate various strong decays widths of the resonance. The growing information around the claimed $Z_c(4025)$, together with the present work, comes to support a $D^* \bar D^*$ molecular state below threshold as an interpretation of the $(D^* \bar D^*)^{\pm}$ peak seen in \cite{besexp}. 

In the present paper we do a thorough investigation of sources of interaction for $D^* \bar D^*$, beginning with the vector exchange, which involves the exchange of heavy vectors. Then we evaluate the interaction from one meson exchange ($\pi,\,\eta ,\, \eta^\prime$) , followed by two pion exchange by a different source of interaction, also involving the $\pi,\,\eta ,\, \eta^\prime$ mesons. We show that these mechanisms are OZI suppressed and because of this we proceed to evaluate contributions from two pion exchange which are not OZI forbidden. Then we evaluate the crossed two pion exchange and also the exchange of two pions which interact among themselves giving rise to a ``$\sigma$'' exchange. We show that even if small, the vector exchange is still the leading source of interaction and obtain a barely bound $D^* \bar D^*$ state close to threshold. We discuss uncertainties in the results and the relationship of the result obtained with the peak observed in \cite{besexp}.

\section{Formalism}
\begin{figure}[htpb]
\centering
\includegraphics[scale=0.58]{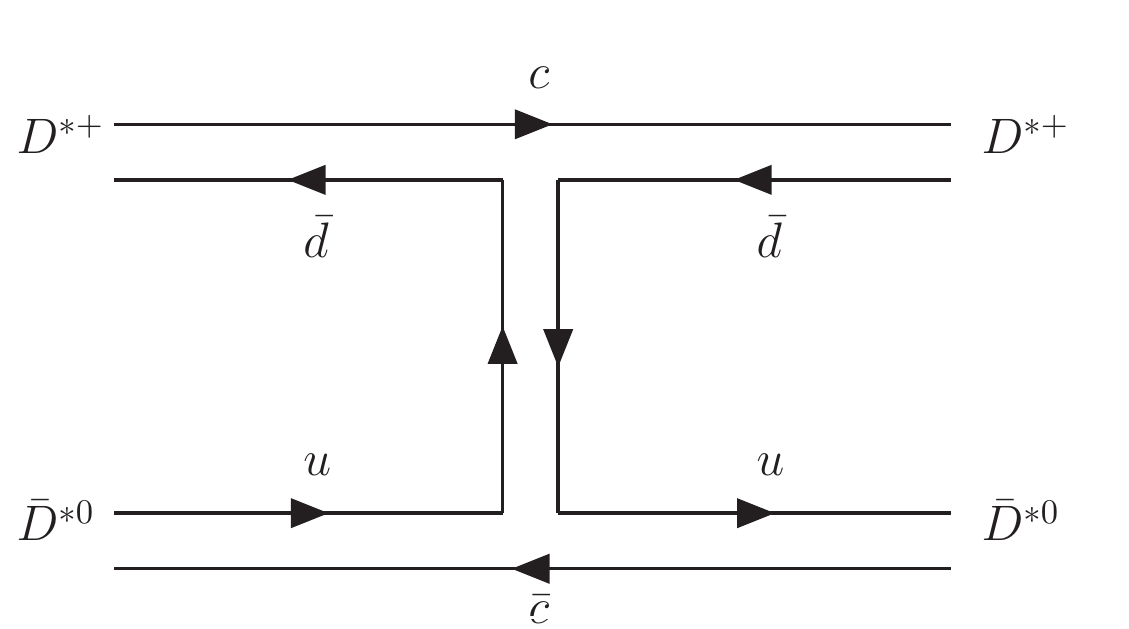}
\caption{Diagrammatic representation of $D^{*+}\bar{D}^{*0}$ interaction exchanging a $q\bar{q}$ and
showing the OZI suppression.}
\label{fig:OZI}
\end{figure}

We want to study states of $I=1$. The first consideration is that one light meson exchange is OZI forbidden. To realize that, we look at the $D^{*+}\bar{D}^{*0}$ interaction diagram in Fig. \ref{fig:OZI} and we see that a $d\bar{d}$ state exchange is forced to be converted into a $u\bar{u }$ state. In terms of physical mesons, that would mean that the $\rho$, $\omega$ cancel if we take equal masses (as it is indeed the case in the hidden gauge approach \cite{raquel2}) and $\pi$, $\eta$, $\eta'$ also cancel if equal masses are taken or for large momenta bigger than the mass of the mesons. We shall show this in detail in subsection \ref{pseudoscalar}.

In view of this cancellation, we will evaluate the contribution of two pion exchange in the next sections, where the OZI restriction no longer holds. Yet we begin by evaluating the contribution of vector exchange that will be the largest one at the end.


\subsection{Vector exchange}
\label{vector}

We follow the approach of Ref. \cite{raquel2}, a study of the vector-vector interaction in the framework of hidden gauge formalism for the channels with quantum numbers $C=0$ and $S=0$. In this paper possible vector-vector states are investigated using a unitary approach in coupled channels.

 The starting point is a Lagrangian coming from the hidden gauge formalism describing the interaction of vector mesons among themselves, 
\begin{equation}
\mathcal{L}=-\frac{1}{4}\langle V_{\mu\nu}V^{\mu\nu}\rangle\ ,
\label{eq:lvv}
\end{equation}
where the symbol $\langle ~ \rangle$ stands for the trace of SU(4), $V_{\mu\nu}$ is defined as
\begin{equation}
V_{\mu\nu}=\partial_{\mu}V_{\nu}-\partial_{\nu}V_{\mu}-ig[V_{\mu},V_{\nu}]\ 
\label{eq:vectensor}
\end{equation}
and $V_{\mu}$ is given by 
\begin{equation}
V_\mu=\left(
\begin{array}{cccc}
\frac{\omega}{\sqrt{2}}+\frac{\rho^0}{\sqrt{2}} & \rho^+ & K^{*+}&\bar{D}^{*0}\\
\rho^- &\frac{\omega}{\sqrt{2}}-\frac{\rho^0}{\sqrt{2}} & K^{*0}&D^{*-}\\
K^{*-} & \bar{K}^{*0} &\phi&D^{*-}_s\\
D^{*0}&D^{*+}&D^{*+}_s&J/\psi
\end{array}
\right)_\mu\ .
\label{eq:vfields}
\end{equation}
The coupling constant is $g=M_{V}/2f_{\pi}$, with $f_{\pi}=93$ MeV the pion decay constant and $M_V\simeq 800$ MeV.

From the Lagrangian in Eq. \eqref{eq:lvv},  two different types of interaction can be derived: a contact interaction, coming from the $[V_{\mu},V_{\nu}]$ term,
\begin{equation}
\mathcal{L}^{(c)}=\frac{g^2}{2}\langle V_{\mu}V_{\nu}V^{\mu}V^{\nu}-V_{\nu}V_{\mu}V^{\mu}V^{\nu}\rangle\ ,
\label{eq:contact}
\end{equation}
and the three-vector vertex
\begin{equation}
\mathcal{L}^{(3V)}=ig\langle(\partial_{\mu}V_{\nu}-\partial_{\nu}V_{\mu})V^{\mu}V^{\nu} \rangle\ .
\label{eq:3v}
\end{equation}
The Lagrangian $\mathcal{L}^{(3V)}$ produces the $VV\rightarrow VV$ interaction by means of the exchange of one vector meson.

The channels we are interested in are the ones with quantum numbers $I=1$, charm $C=0$ and strangeness $S=0$, which are $D^*\bar{D}^*, K^*\bar{K}^*, \rho\rho, \rho\omega, \rho J/\psi, \rho\phi$. From the Lagrangians in Eqs. \eqref{eq:contact} and \eqref{eq:3v}, the amplitudes that will be used as the kernel to solve the Bethe-Salpeter equation can be evaluated. The reader can find all the details of the calculation in Ref. \cite{raquel2}.

Here, we will only consider the case with $J=2$ since this is the only spin channel where the interaction gives an attractive potential for $D^*\bar{D}^*\rightarrow D^*\bar{D}^*$. In \cite{raquel2},  in addition to the $\rho J/\psi$ channel, which is the most important after the $D^*\bar{D}^*$, the $\rho\rho$, $\rho\omega$, $\rho\phi$ light vector channels were also considered. However, the thresholds of these channels are situated at energies much smaller than the mass of the state we are looking for, such that the results would be only slightly affected by their inclusion.

The $\rho J/\psi$ channel plays an important role in this problem. Indeed, the transition potential of $D^*\bar{D}^*\rightarrow \rho J/\psi$ has a strength almost four times bigger than the $D^*\bar{D}^*\rightarrow D^*\bar{D}^*$ transition.

The expressions of these potentials are reported in the following equations, including both the contact and the vector-exchange term:
\begin{equation}
t_{D^{*}\bar{D}^{*}\rightarrow D^{*}\bar{D}^{*}}=-g_{D}^2+g_D^2\frac{(2m_{\omega}^2m_{\rho}^2+m_{J/\psi}^2(-m_{\omega}^2+m_{\rho}^2))(4m_{D^*}^2-3s)}{4m_{J/\psi}^2 m_{\omega}^2 m_{\rho}^2}\ ,
\label{eq:DDDD}
\end{equation} 
\begin{equation}
t_{D^{*}\bar{D}^{*}\rightarrow \rho J/\psi}=-2gg_{D}+gg_D\frac{2m_{D^*}^2+m_{J/\psi}^2+m_{\rho}^2-3s}{m_{D^*}^2}\ ,
\label{eq:DDRJ}
\end{equation} 
where $m_{\rho}$, $m_{\omega}$ and $m_{J/\psi}$ are the masses of the $\rho$, $\omega$ and $J/\psi$ mesons respectively. The constant $g_D=m_{D^*}/(2f_D)$, which was used in \cite{raquel2}, is analogous to the coupling $g$ for light mesons, with $f_D=206/\sqrt{2}=145.66$ MeV. However, as we discuss below, we can use constrains of heavy quark spin symmetry to provide a more accurate coupling.

The exercise to relate the $D^*D\pi$ vertex to the $K^*K\pi$ in \cite{xiaoliang} is repeated  in that work for the Weinberg-Tomozawa term that we are considering now, based on the exchange of vector mesons. The $D^*\bar{D}^*\rightarrow D^*\bar{D}^*$ is now mediated  by $J/\psi$ exchange ($c\bar{c}$) in analogy to the $\phi$ exchange in  $K^*\bar{K}^*\rightarrow K^*\bar{K}^*$. The rules of heavy quark spin symmetry, $HQSS$ \cite{wise}, can be obtained from the impulse approximation at the quark level assuming the $s$ and $c$ as spectators. Then, given the $(2\omega)^{-1/2}$ normalization factors of the fields at the meson level, there is a factor $\omega_{D^*}/\omega_{K^*}$ between the $D^*D^*J/\psi$ and the $K^*K^*\phi$ vertices. Since the $K^*K^*\phi$ vertex is proportional to $\omega_ {K^*}$, the  $D^*D^*J/\psi$ will have the same proportionality coefficient multiplied by $\omega_{D^*}$, which is what the straight application of $SU(4)$ provides in this case. Note that the vector exchange term in Eq. \eqref{eq:DDDD} at the $D^*\bar{D}^*$ threshold, for simplicity, gives, with $m_{\omega}=m_{\rho}$, $g_D^2\frac{m_{D^*}^2}{m_{J/\psi}^2}$. There we see explicitly the energy of the $m_{D^*}$ from the two vertices and $m_{J/\psi}^2$ from the $J/\psi$ propagator. According with the previous argument this should be $g^2\frac{m_{D^*}^2}{m_{J/\psi}^2}$. We, thus, use here the normal $g$ coupling which is in agreement with the heavy quark spin symmetry . For consistency, we also take $g^2$ in the contact term, which is smaller than the  $J/\psi$ exchange one, and in the  transition potential of Eq. \eqref{eq:DDRJ}. The use of the new coupling will have as a consequence the reduction of the binding of the $I=1$ state with respect to the one found in \cite{raquel2}. 

\begin{figure}
  \centering
  \subfigure[]{\label{fig:tDD}\includegraphics[width=0.4965\textwidth]{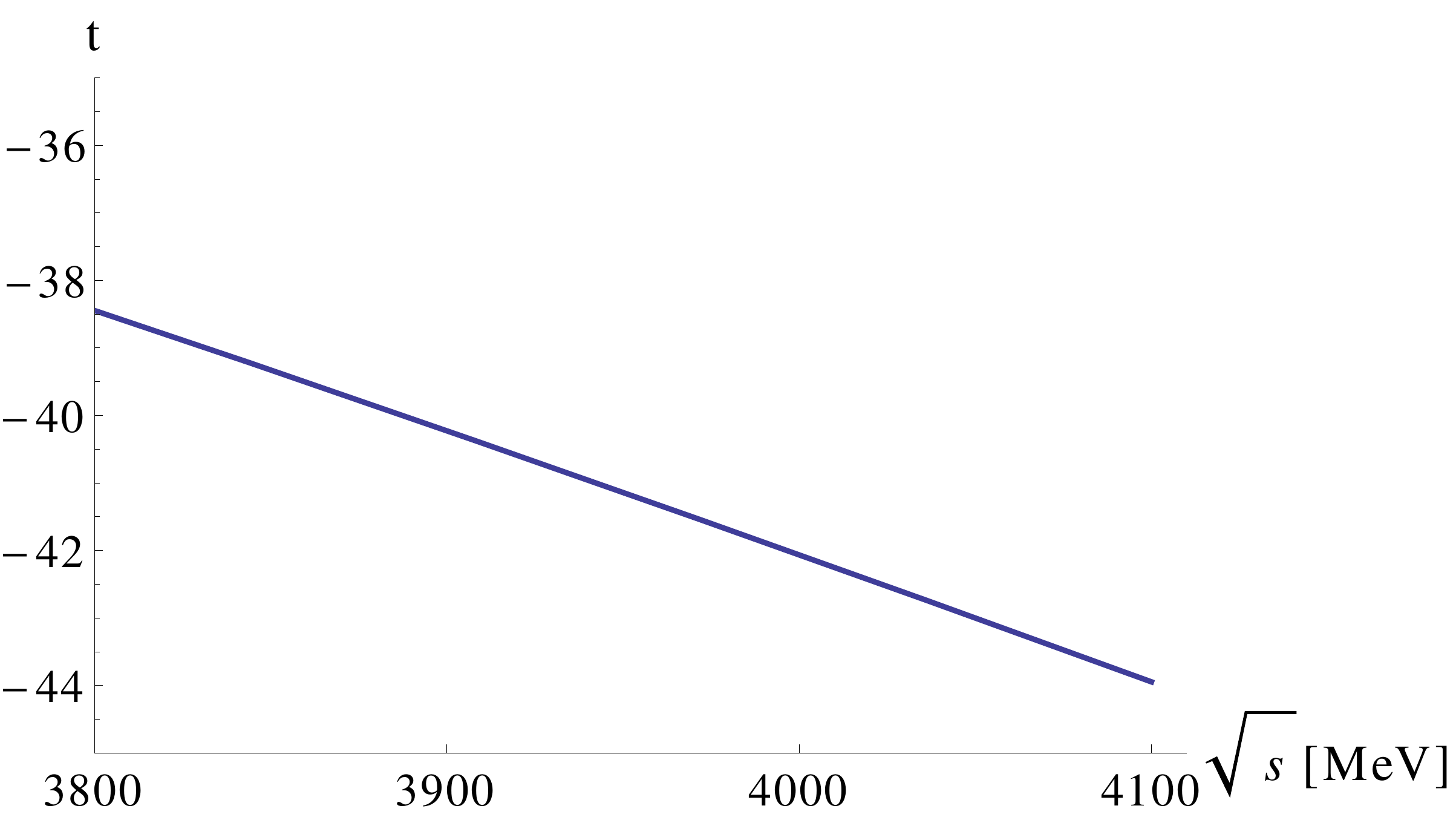}}                
   \subfigure[]{\label{fig:tRJ}\includegraphics[width=0.4965\textwidth]{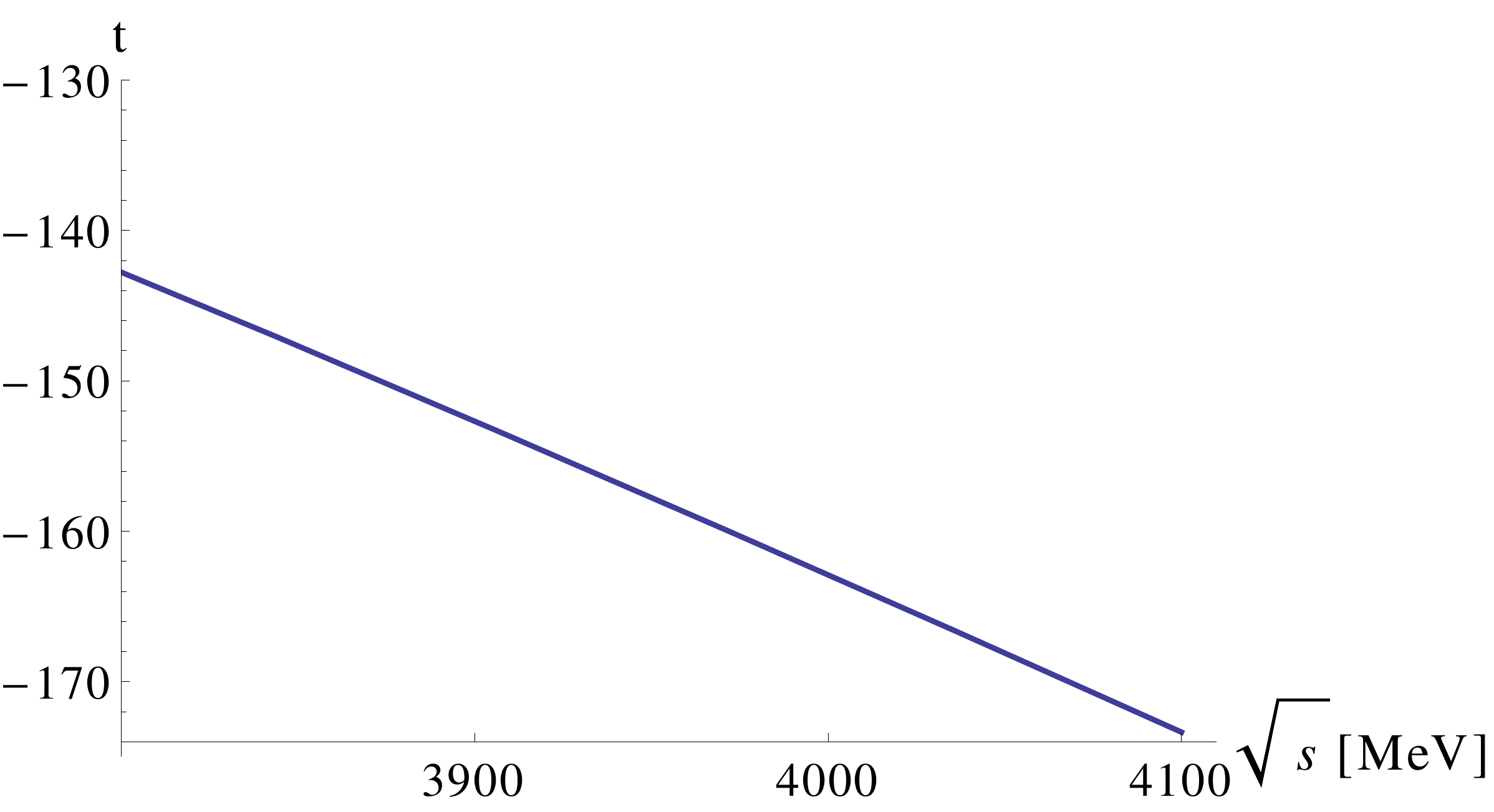}}
  \caption{Potentials $t_{D^{*}\bar{D}^{*}\rightarrow D^{*}\bar{D}^{*}}$  $(a)$ and  $t_{D^{*}\bar{D}^{*}\rightarrow \rho J/\psi}$ $(b)$ as functions of the center of mass energy $\sqrt{s}$.}
\label{fig:vex}
\end{figure}

The two potentials are plotted in Fig. \ref{fig:vex} as functions of the centre of mass energy $\sqrt{s}$.
The expressions of Eqs. \eqref{eq:DDDD}  and \eqref{eq:DDRJ} provide the potential $V$ that must be used to solve the Bethe-Salpeter equation in coupled channels
\begin{equation}
T=(1-VG)^{-1}V\ ,
\label{eq:BS}
\end{equation}
where $V$ is a $2\times2$ matrix with elements $V_{11}=t_{D^{*}\bar{D}^{*}\rightarrow D^{*}\bar{D}^{*}}$, $V_{12}=V_{21}=t_{D^{*}\bar{D}^{*}\rightarrow \rho J/\psi}$, and $V_{22}=0$. The matrix $G$ is the $2\times 2$ diagonal loop function matrix whose elements are given by
\begin{equation}
G_i=i\int\frac{d^4q}{(2\pi)^4}\frac{1}{q^2-m_1^2+i\epsilon}\frac{1}{(q-P)^2-m_2^2+i\epsilon}\ ,
\label{eq:loopex}
\end{equation}
with  $m_1$ and $m_2$ the masses of the two mesons involved in the loop in the channel $i$ and $P$ the total four-momentum of the mesons.

 After the integration in $dq^0$,  Eq. \eqref{eq:loopex} becomes
\begin{equation}
\label{eq:1striemann}
G_i=\int\frac{d^3q}{(2\pi)^3}\,\frac{\omega_1+\omega_2}{2\omega_1\omega_2}\,\frac{1}{(P^0)^2-(\omega_1+\omega_2)^2+i\epsilon}\ ,
\end{equation}
which is regularized by means of a cutoff in the three-momentum $q_{max}$.

The function $G_i$ can be also written in dimensional regularization as
\begin{equation}
\begin{split}
\label{eq:loopexdm}
G_i&=\frac{1}{16\pi^2}(\alpha_i+\log\frac{m_1^2}{\mu^2}+\frac{m_2^2-m_1^2+s}{2s}\log\frac{m_2^2}{m_1^2}+\frac{p}{\sqrt{s}}(\log\frac{s-m_2^2+m_1^2+2p\sqrt{s}}{-s+m_2^2-m_1^2+2p\sqrt{s}}\\&+\log\frac{s+m_2^2-m_1^2+2p\sqrt{s}}{-s-m_2^2+m_1^2+2p\sqrt{s}}))\ ,
\end{split}
\end{equation}
where $p$ is the three-momentum of the mesons in the centre of mass
\begin{equation}
p=\frac{\sqrt{(s-(m_1+m_2)^2)(s-(m_1-m_2)^2)}}{2\sqrt{s}}\ .
\end{equation}


\subsection{The $D^*\bar{D}^*$ interaction via light pseudoscalar exchange}
\label{pseudoscalar}

\begin{figure}[htpb]
\centering
\includegraphics[scale=0.7]{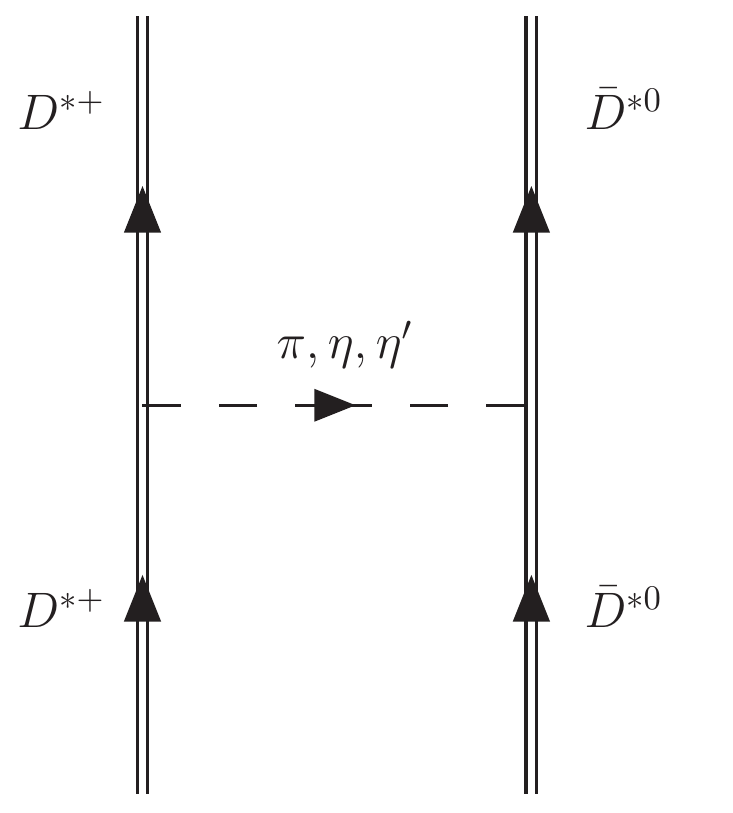}
\caption{Diagrammatic representation of $D^*\bar{D}^*$ interaction via light pseudoscalar exchange.}
\label{fig:light}
\end{figure}

The pseudoscalar exchange between vector mesons, shown in Fig. \ref{fig:light}, proceeds via the anomalous 
vector-vector-pseudoscalar ($VVP$) coupling,
\begin{equation}
\mathcal{L}=\frac{G}{\sqrt{2}}\epsilon^{\mu \nu \alpha \beta}\langle \partial_{\mu}V_{\nu} \partial_{\alpha}V_{\beta}P \rangle\ ,
\label{eq:vvp}
\end{equation}
where $G=3M_{V}^{2}/16 \pi^{2}f_{\pi}^{3}$, with $M_{V}\simeq 800$ MeV, $f_{\pi}=93$ MeV. The matrix $V$ is given by Eq. \eqref{eq:vfields}, while $P$ contains the 15-plet of the pseudoscalar mesons written in the physical basis in which $\eta$, $\eta'$ mixing is considered \cite{gamphi3770}, 
\begin{equation}
P=\left(
\begin{array}{cccc}
\frac{\eta}{\sqrt{3}}+\frac{\eta'}{\sqrt{6}}+\frac{\pi^0}{\sqrt{2}} & \pi^+ & K^+&\bar{D}^0\\
\pi^- &\frac{\eta}{\sqrt{3}}+\frac{\eta'}{\sqrt{6}}-\frac{\pi^0}{\sqrt{2}} & K^{0}&D^-\\
K^{-} & \bar{K}^{0} &-\frac{\eta}{\sqrt{3}}+\sqrt{\frac{2}{3}}\eta'&D^-_s\\
D^0&D^+&D^+_s&\eta_c
\end{array}
\right)\ .
\label{eq:pfields}
\end{equation}

It is easy to evaluate the contribution of the pseudoscalar exchange and, close to $D^*\bar{D}^*$ 
threshold, we find
\begin{equation}
t \simeq -\frac{G^2}{2}m_{D^*}^{2} ~ \vec{q} \cdot (\vec{\epsilon}_{1} \times \vec{\epsilon}_{3})~\vec{q} \cdot (\vec{\epsilon}_{2} \times \vec{\epsilon}_{4}) \Big( -\frac{1}{2} \frac{1}{q^{2}-m_{\pi}^2} +\frac{1}{3}\frac{1}{q^{2}-m_{\eta}^2}+\frac{1}{6}\frac{1}{q^{2}-m_{\eta'}^2}\Big)\ ,
\label{eq:tanomalous}
\end{equation}
where $\vec{\epsilon}_{1}$, $\vec{\epsilon}_{2}$ stand for the initial polarizations of the vector mesons, $\vec{\epsilon}_{1}$, $\vec{\epsilon}_{3}$ for the final ones and $\vec{q}$ is the momentum transfer.

Note again that $t$ in Eq. \eqref{eq:tanomalous} is already proportional to $\omega_{D^*}^2$ ($m_{D^*}^2$ at threshold) and the factor proportional to $\omega_D^*$ in each vertex demanded in $HQSS$ is automatically  included in Eq. \eqref{eq:tanomalous} as it was also the case in the Weinberg-Tomozawa terms. 

Since we are concerned in s-waves, we can take $\vec{q}_{i}\vec{q}_{j}\rightarrow 
\frac{1}{3}\vec{q}^{\ 2}\delta_{ij}$, which leads to the spin structure
\begin{equation}
\label{eq:order}
(\vec{\epsilon_{1}} \times \vec{\epsilon_{3}}) (\vec{\epsilon_{2}} \times \vec{\epsilon_{4}})=\epsilon_{i}
\epsilon_{i}\epsilon_{j}\epsilon_{j} - \epsilon_{i}\epsilon_{j}\epsilon_{j}\epsilon_{i}\ ,
\end{equation}
where the order of the polarization vectors in the right-hand side of Eq. \eqref{eq:order} is 1, 2, 3 and 4. By recalling the form of the spin projector operators  \cite{raquel},
\begin{equation}
\begin{split}
&\mathcal{P}^{(0)}=\frac{1}{3}\epsilon_{i}\epsilon_{i}\epsilon_{j}\epsilon_{j}\ ,\\
&\mathcal{P}^{(1)}=\frac{1}{2}\left(\epsilon_{i}\epsilon_{j}\epsilon_{i}\epsilon_{j}-\epsilon_{i}\epsilon_{j}\epsilon_{j}\epsilon_{i}\right)\ ,\\
&\mathcal{P}^{(2)}=\frac{1}{2}\left(\epsilon_{i}\epsilon_{j}\epsilon_{i}\epsilon_{j}+\epsilon_{i}\epsilon_{j}\epsilon_{j}\epsilon_{i}\right)-\frac{1}{3}\epsilon_{i}\epsilon_{i}\epsilon_{j}\epsilon_{j}\ ,
\label{eq:projection}
\end{split}
\end{equation}
we see that
\begin{equation}
(\vec{\epsilon_{1}} \times \vec{\epsilon_{3}}) (\vec{\epsilon_{2}} \times \vec{\epsilon_{4}})=2\mathcal{P}^{(0)}+
\mathcal{P}^{(1)}-\mathcal{P}^{(2)}\ .
\end{equation}

We are interested in the spin $J=2$ component and thus we have
\begin{equation}
t^{(2)} \simeq\frac{G^2}{2}m_{D^*}^{2} ~ \vec{q}^{\ 2} \Big( \frac{1}{2} \frac{1}{\vec{q}^{\ 2}+m_{\pi}^2} -\frac{1}{3}\frac{1}{\vec{q}^{\ 2}+m_{\eta}^2}-\frac{1}{6}\frac{1}{\vec{q}^{\ 2}+m_{\eta'}^2}\Big)\Big( \frac{\Lambda^2}{\Lambda^2+\vec{q}^{\ 2}} \Big)\ ,
\label{eq:tspin2}
\end{equation}
where we have taken into account that $q^{0}=0$ and we have introduced a customary convergence form factor $\frac{\Lambda^2}{\Lambda^2+\vec{q}^{\ 2}}$, with $\Lambda=1000$ MeV \cite{toki}. 

\begin{figure}[htpb]
\centering
\includegraphics[scale=0.35]{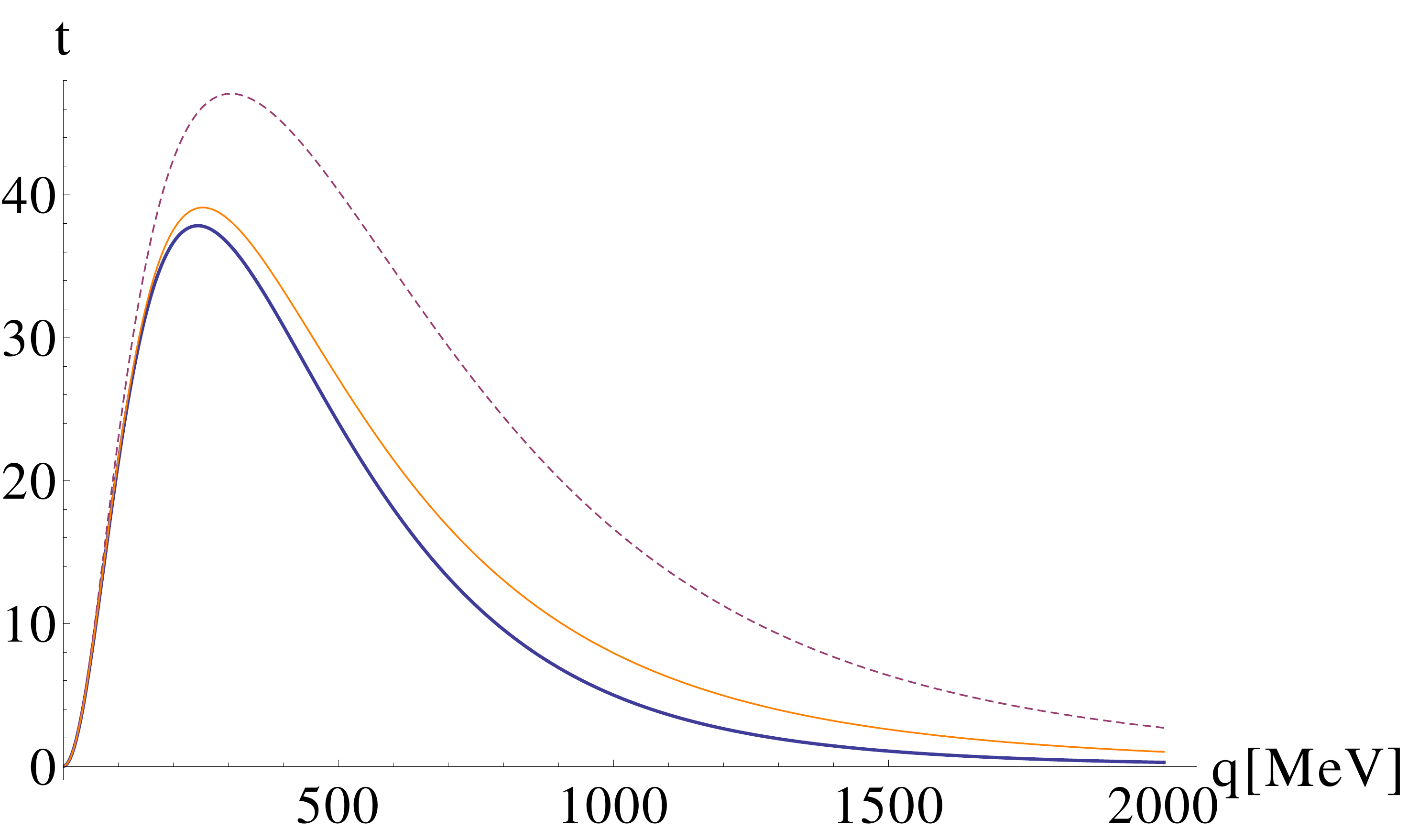}
\caption{Potential $t^{(2)}$ for the exchange of one light meson ($\pi$ plus $\eta$ plus $\eta^{\prime}$, thick line), one pion (dashed line) and $\pi$ plus $\eta$ (thin line) as a function of the momentum transferred in the process.}
\label{fig:tlightex}
\end{figure}

\begin{figure}[htpb]
\centering
\includegraphics[scale=0.5]{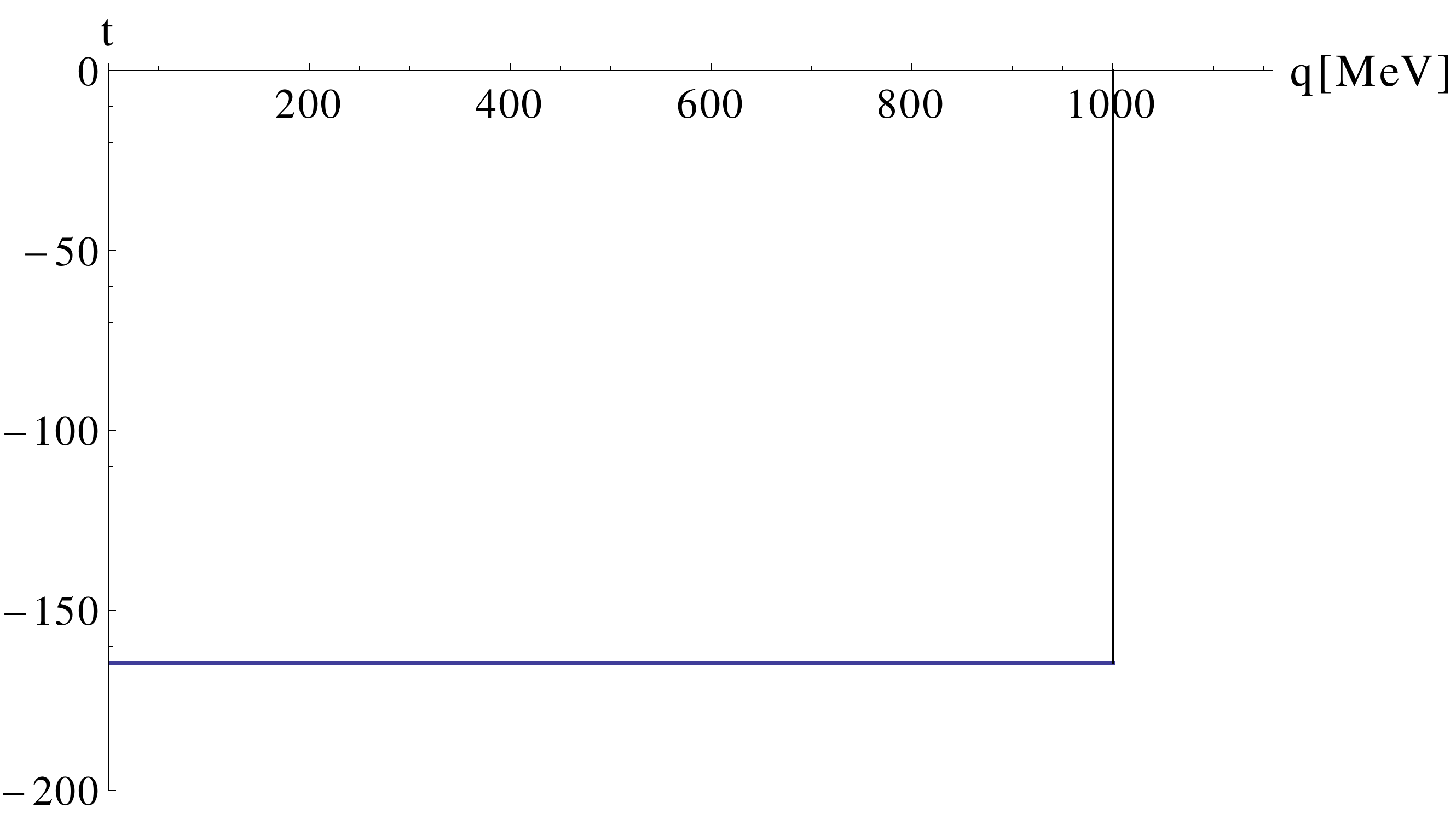}
\caption{$t_{D^{*}\bar{D}^{*}\rightarrow \rho J/\psi}$ as a function of the momentum transferred in the process $\vec{q}$ (with $\vec{p}=0$).}
\label{fig:vecex}
\end{figure}

In Eq. (\ref{eq:tspin2}) we observe the explicit cancellation of the exchange of $\pi$, $\eta$, $\eta'$ 
in the limit of equal masses. In Fig. \ref{fig:tlightex} we plot $t^{(2)}$ as a function of $\vec{q}$ in order to compare it with the vector exchange potential for the $D^{*}\bar{D}^{*}\rightarrow \rho J/\psi$ transition, shown in Fig. \ref{fig:vecex}. In Fig. \ref{fig:tlightex} we can see explicitly the cancellation between $\pi,\, \eta$ and $\eta^\prime$. On the other hand, the amplitude is proportional to $\vec{q}^{\ 2}$ and this is a very small quantity around threshold, where the states reported here are found. But this argument is only relevant for the tree level amplitude and in the solution of the Bethe-Salpeter equation we shall have loops which involve larger $\vec{q}$. The small values of $t^{(2)}$ at small $\vec{q}$, where one pion exchange is clearly dominant, and the cancellations at large values of $\vec{q}$, render this term small in all the range of $\vec{q}$. The results of Fig. \ref{fig:tlightex} are shown for an energy of $D^*\bar{D}^*$ at threshold and do not change appreciably in the range of energies where we are concerned.

According to \cite{daniwf} the vector exchange potential, together with the cutoff in $G$, can be reinterpreted as a separable potential of the type $V(\vec{p},\vec{p}^{\ \prime})=V\theta(p_{max}-|\vec{p}\,|)\theta(p_{max}-|\vec{p}^{\ \prime}|)$. Although this is not a function of $\vec{q}=\vec{p}-\vec{p}^{\ \prime}$, as in Eq. \eqref{eq:tspin2}, if we take $\vec{p}=0$ and vary $\vec{p}^{\ \prime}$ in a loop, then $\vec{p}^{\ \prime}$ behaves as $\vec{q}$ and this allows a fair comparison Figs. \ref{fig:tlightex} and \ref{fig:vecex}.

We can see that the strength of the $D^*\bar D^*\rightarrow \rho\, J/\psi$ transition due to the vector exchange is much larger than the one pion exchange potential in all the range of $q$. If we integrate $\int d^{3}q V(q)$ up to $q=1000$ MeV in both cases, we find the integral ten times larger in the $D^*\bar D^*\rightarrow \rho\, J/\psi$ transition case, and we neglect the contribution of one meson exchange in our calculation.

\subsection{Iterated two meson exchange}
\label{iterated}

\begin{figure}[htpb]
\centering
\includegraphics[scale=0.7]{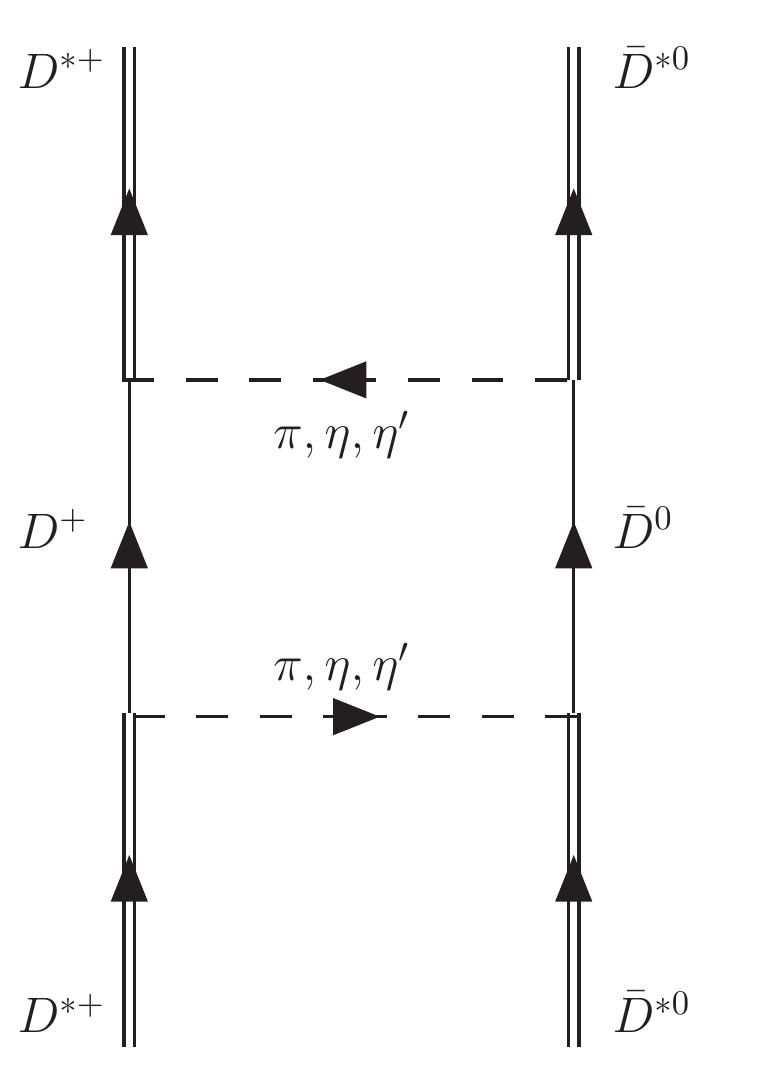}
\caption{Iterated pseudocalar exchange with intermediate $D\bar{D}$ states.}
\label{fig:box}
\end{figure}

In the former section we saw that the exchange of $\pi,\,\eta, \, \eta'$ between the vectors gave rise to strong cancellations and was small compared to the typical contribution of vector exchange.

Here we show that there is a different kind of vector exchange involving this time the $VPP$ vertex, instead of the anomalous $VVP$ one that we have studied before. However, the $VV\, \rightarrow VV$ transition needs a box diagram to accommodate the intermediate $PP$ states. This is depicted in Fig. \ref{fig:box}. 

One needs now the $VPP$ Lagrangian given by 
\begin{equation}
\label{eq:lagrangianhg}
\mathcal{L}_{PPV}=-ig\ \langle V^\mu [P,\partial_\mu P]\rangle\ ,
\end{equation}
where the constant $g$ is the strong coupling of the $D^*$ meson to $D\pi$. With the vector at rest, as we consider for the evaluation, Eq. \eqref{eq:lagrangianhg} provides vertices of the type $g\vec{\epsilon}\cdot\vec{p}$ (since $\epsilon^0=0$). We use for $g$ the theoretical value obtained  within the heavy quark spin symmetry approach, which is given by the $SU(3)$ value of $g=\frac{m_V}{2f}$, multiplied by $m_{D^*}/m_{K^*}$ \cite{xiaoliang}. Note that in this case the vertex is not proportional to $\omega_{D^*}$ and hence the factor $\omega_{D^*}/\omega_{K^*}$ (equal to $m_{D^*}/m_{K^*}$ at threshold) remains. This gives the effective value of $\tilde{g}=9.40$. With this value, we obtain $71$ KeV for the width of $D^{*+}\rightarrow D^0\pi^+$ compared to the experimental value $(65\pm15)$ KeV of \cite{cleo2}.

The amplitude for the box diagram of Fig. \ref{fig:box} is given by
\begin{eqnarray}
t &=& i\tilde{g}^{4}\int \frac{d^{4}p}{(2\pi)^{4}}\, \vec{\epsilon}_{1}\cdot 2\vec{p} \,\vec{\epsilon}_{2}\cdot 2\vec{p}\, \vec{\epsilon}_{3}\cdot 2\vec{p}\, \vec{\epsilon}_{4}\cdot 2\vec{p} \,\Big( -\frac{1}{2} \frac{1}{\vec{p}^{\ 2}-m_{\pi}^2+i\epsilon} +\frac{1}{3}\frac{1}{\vec{p}^{\ 2}-m_{\eta}^2+i\epsilon}\nonumber \\
&+&\frac{1}{6}\frac{1}{\vec{p}^{\ 2}-m_{\eta'}^2+i\epsilon}\Big)^2 \, \frac{1}{(2E_{D}(\vec{p}\,))^{2}}\,\frac{1}{m_{D^*}-p^{0}-E_{D}(\vec{p}\,)+i\epsilon}\,\frac{1}{m_{D^*}-p^{0}-E_{D}(\vec{p}\,)+i\epsilon}\ ,
\label{eq:box1}
\end{eqnarray}
where $E_{D}(\vec{p}\,)=\sqrt{\vec{p}^{\ 2}+m_{D}^2}$ is the energy of the $D$ meson.

By symmetry reasons we can substitute
\begin{equation}
p_{i}p_{j}p_{k}p_{m}\, \rightarrow \frac{1}{15}(\delta_{ij}\delta_{km}+\delta_{ik}\delta_{jm}+\delta_{im}\delta_{jk})\, \vec{p}^{\ 4}\ ,
\label{eq:pppp}
\end{equation}
which renders the spin combination into
\begin{equation}
\frac{1}{15}(\epsilon_{1i}\epsilon_{2i}\epsilon_{3m}\epsilon_{4m}+\epsilon_{1i}\epsilon_{2j}\epsilon_{3i}\epsilon_{4i}+
\epsilon_{1i}\epsilon_{2j}\epsilon_{3j}\epsilon_{4i})\
\label{eq:spinbox} .
\end{equation}
Taking into account the spin projections of Eqs. \eqref{eq:projection}, the combination of spin that we have in Eq. \eqref{eq:spinbox} is 
\begin{equation}
\frac{1}{15}(5\mathcal{P}^{(0)}+2\mathcal{P}^{(2)})\ .
\label{eq:spinbox2}
\end{equation}
Then, performing analytically the $p^{0}$ integration in Eq. \eqref{eq:box1}, we obtain
\begin{equation}
\tilde{t}=\frac{1}{4}t_{\pi\pi}+\frac{1}{9}t_{\eta\eta}+\frac{1}{36}t_{\eta'\eta'}-\frac{1}{3}t_{\pi\eta}-\frac{1}{6}t_{\pi\eta'}+\frac{1}{9}t_{\eta\eta'}\ ,
\label{eq:box2}
\end{equation}
where 
\begin{eqnarray}
t_{12}&=&\tilde{g}^{4}S_{J}\int \frac{d^{3}p}{(2\pi)^{3}}\,\vec{p}^{\ 4}\Big( \frac{\Lambda^2}{\Lambda^2+\vec{p}^{\ 2}}\Big)^{4}\frac{1}{m_{D^*}+\omega_{1}-E_{D}(\vec{p})\pm i\epsilon}\,\frac{1}{m_{D^*}+\omega_{2}-E_{D}(\vec{p})\pm i\epsilon}\nonumber \\
&\times&\frac{1}{(E_{D}(\vec{p}\,))^2}\Big (\frac{1}{2\omega_1\omega_2}\,\frac{1}{\omega_1+\omega_2}\,\frac{\textrm{Num}}{m_{D^{*}}-\omega_1-E_{D}(\vec{p}\,)+i\epsilon}\,\frac{1}{m_{D^{*}}-\omega_2-E_{D}(\vec{p}\,)+i\epsilon} \nonumber \\
&+&\frac{1}{E_D(\vec{p}\,)-m_{D^*}+\omega_1+i\epsilon}\,\frac{1}{E_D(\vec{p}\,)-m_{D^*}+\omega_2+i\epsilon}\,\frac{1}{2m_{D^*}-2E_{D}(\vec{p}\,)+i\epsilon}\Big )\ ,
\label{eq:t12}
\end{eqnarray}
where the subscript $12$ stands for the two light mesons exchanged, $\omega_1$ and $\omega_2$ are their energies, 
\begin{equation}
S_J=\begin{cases} 
\frac{4}{3}\ \ \ \ \ \ J=0\\ 
\\
\frac{8}{15}\ \ \ \ \ \ J=2\ ,
\end{cases}
\end{equation}
and 
\begin{equation}
\textrm{Num}=-(\omega_1^2+\omega_2^2+\omega_1\omega_2)+(m_{D^*}-E_{D}(\vec{p}\,))^2\ .
\end{equation}

The former calculation has been done at threshold. The momentum transfer dependence on $\vec{q}$ can be obtained easily from Eq. \eqref{eq:t12} by taking for the initial and final states four-momenta $p_1=(p_1^0,\vec{q}/2)$, $p_2=(p_2^0,-\vec{q}/2)$, $p_3=(p_3^0,-\vec{q}/2)$ and $p_4=(p_4^0,\vec{q}/2)$.

\begin{figure}[htpb]
\centering
\includegraphics[scale=0.35]{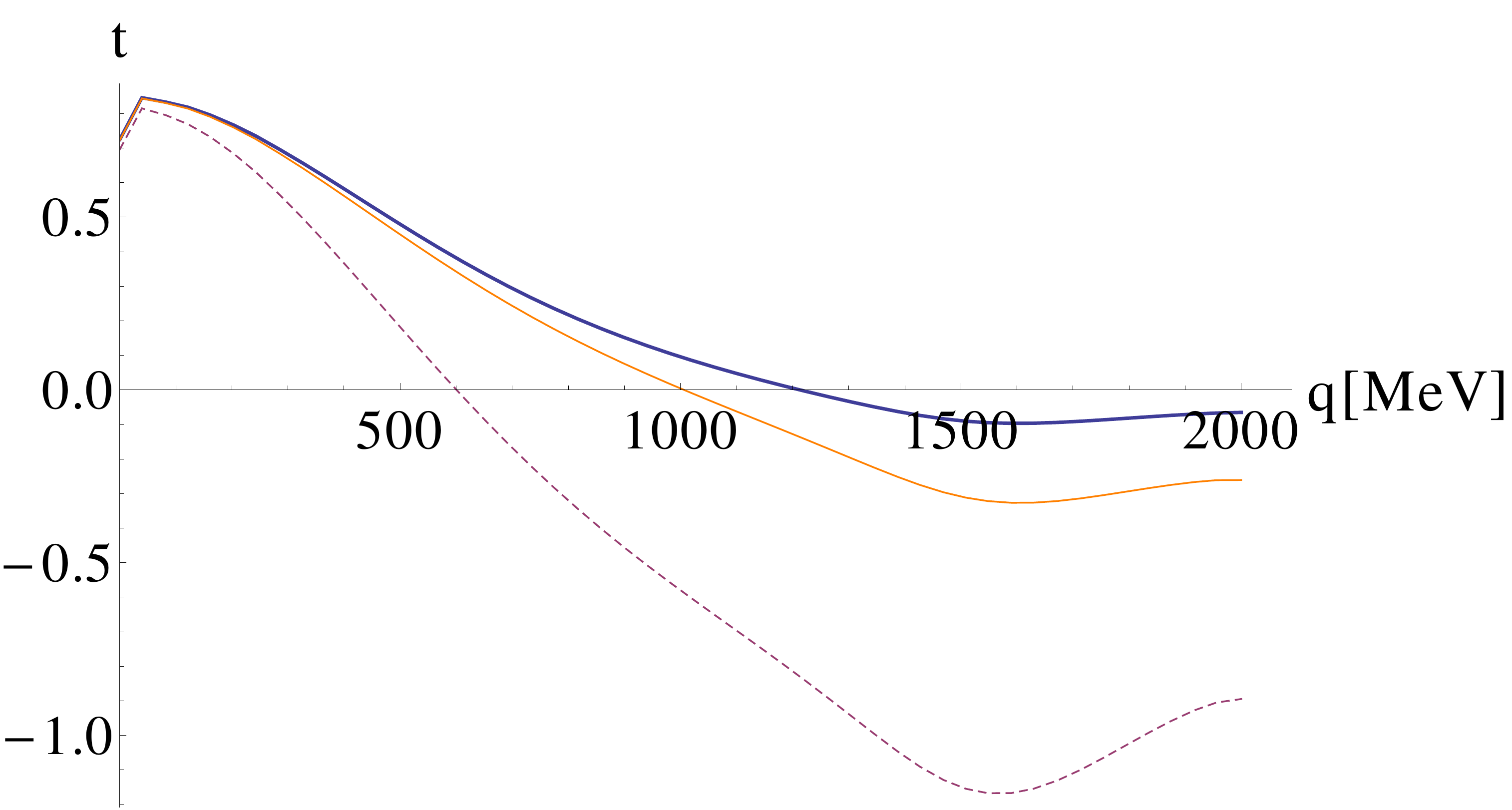}
\caption{Amplitudes $\tilde{t}$ (thick line), accounting for $\pi$ plus $\eta$ plus $\eta^{\prime}$ exchange, $\frac{t_{\pi\pi}}{4}$ (dashed line), accounting for only pion exchange, and $\frac{t_{\pi\pi}}{4}+\frac{t_{\eta\eta}}{9}-\frac{t_{\pi\eta}}{3}$ (thin line), accounting for $\pi$ plus $\eta$ exchange as functions of the transferred momentum.}
\label{fig:box2}
\end{figure}

In Fig. \ref{fig:box2} we show the results, for $J=2$, of $\tilde{t}$ compared with $\frac{t_{\pi\pi}}{4}$ and $\frac{t_{\pi\pi}}{4}+\frac{t_{\eta\eta}}{9}-\frac{t_{\pi\eta}}{3}$. There is a cancellation between $\pi$, $\eta$ and $\eta'$ as in the anomalous exchange, which is exact in the limit of equal masses for the mesons. Once again, we can see that the contribution is much smaller than the typical term due to vector exchange.

\subsection{The $D^*\bar{D}^*$ interaction by means of $\sigma$ exchange}
\label{sigma}

The nucleon-nucleon interaction calls for an intermediate range attraction which was traditionally taken into account by means of ``$\sigma$" exchange. With ups and downs the $\sigma$ resonance appears now in the PDG \cite{pdg} as the $f_0(500)$. This resonance appears unavoidably in a study of the $\pi \pi$ interaction with a unitary approach using as input the kernel from the chiral Lagrangians \cite{oller,kaiser,rios}. The analysis of $\pi \pi$ data with Roy equations allows one to establish the mass and width of this resonance with some precision \cite{colangelo,josekaminski}, compatible with the prediction of the chiral unitary approach, with mass around 460 MeV and half width around 280 MeV. From this point of view it was interesting to provide a microscopic picture for $\sigma$ exchange, based on the nature of the $\sigma$ resonance stemming from the interaction of two pions. This job was done in \cite{toki} considering the exchange of two correlated (interacting) pions for the NN interaction. In this section we extend these ideas to the interaction of $D^* \bar D^*$. 

\begin{figure}[htpb]
\centering
\includegraphics[scale=0.57]{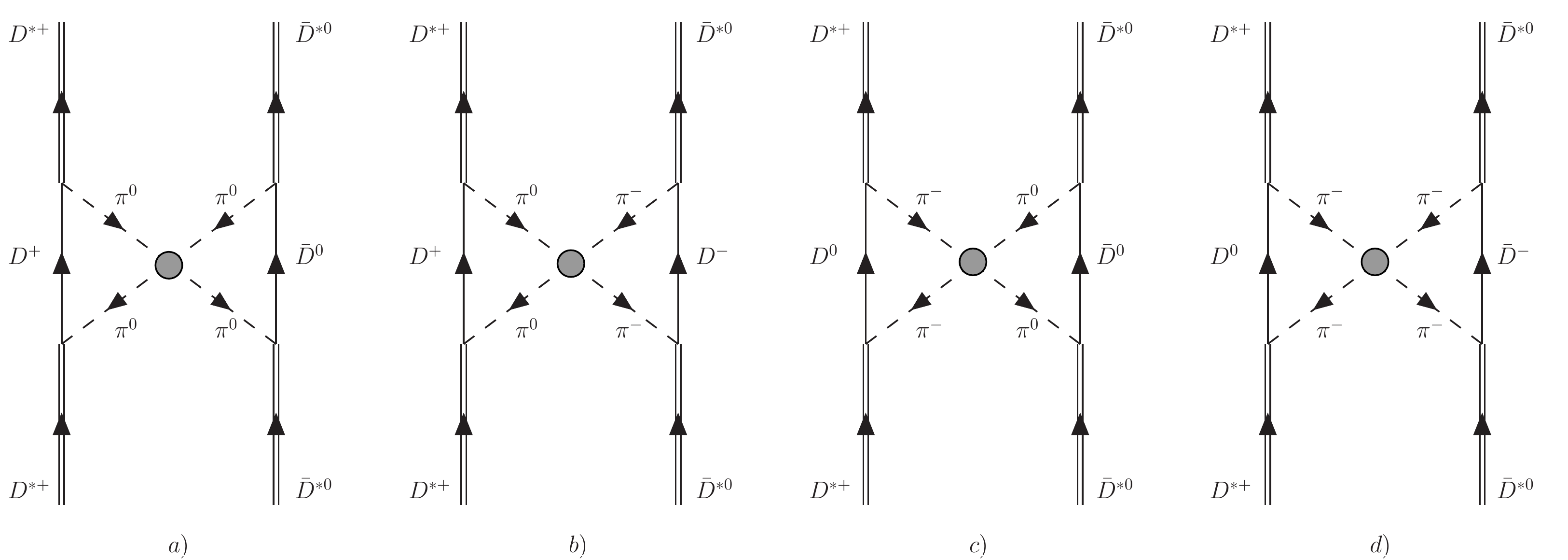}
\caption{Lowest order $\pi\pi$ interaction in the $I=1$ channel for $D^*\bar{D}^*\rightarrow D^*\bar{D}^*$.}
\label{fig:sigmadiag}
\end{figure}

We have four diagrams contributing to this process and they are shown in Fig. \ref{fig:sigmadiag}. Each one of  them contains four $PPV$ vertices involving a $D^*$ ($\bar{D}^*$) vector meson and the two pseudoscalar, the pion and the $D$ ($\bar{D}$) meson. Their evaluation is easily done with the local hidden gauge Lagrangians  \cite{hidden1, hidden2, hidden3, hidden4,roca}, which are very useful when dealing with vector mesons. The grey circle in the crossing of the pion lines indicates that we have there the $\pi\pi$ scattering amplitude.


The Lagrangian we need in order to evaluate the amplitudes of the diagrams in Fig. \ref{fig:sigmadiag} is the one in Eq. \ref{eq:lagrangianhg}. Using it we can write the 
vertices as
\begin{equation}
-it_ {PPV}=-ig\,C(p_D+p_{\pi})_{\mu}\epsilon_V^{\mu}\ ,
\label{eq:ampl_vert}
\end{equation}
where $p_D$ and $p_{\pi}$ are the four-momenta of the $D$ meson and of the pion, respectively, and $\epsilon_V$ is the polarization vector of the $D^*$ meson in the vertex.


The amplitude of the process can be written as
\begin{equation}
-it_{\sigma}=-i\ V^2(\frac{1}{4}t_{\pi^0\pi^0\rightarrow\pi^0\pi^0}+\frac{1}{2}t_{\pi^0\pi^0\rightarrow\pi^+\pi^-}+\frac{1}{2}t_{\pi^+\pi^-\rightarrow\pi^0\pi^0}+t_{\pi^+\pi^-\rightarrow\pi^+\pi^-})\ ,
\label{eq:ampl}
\end{equation}
where the factor $V$ is the contribution to the diagram of the triangular loops, which we shall evaluate later.  Note that, in order to write the amplitude, we must assume two initial pions and two final pions all pointing to the right in the diagrams of Fig. \ref{fig:sigmadiag}, hence providing the amplitude of Eq. \eqref{eq:ampl}.

Considering the unitary normalization of the $\pi\pi$ states \cite{oller}, 
\begin{equation}
|\pi\pi, I=0\rangle = -\frac{1}{\sqrt{6}}|\pi^0\pi^0+\pi^+\pi^-+\pi^-\pi^+\rangle\ ,
\label{eq:norm}
\end{equation}
and writing explicitly the isoscalar amplitude
\begin{equation}
t^{I=0}_{\pi\pi\rightarrow\pi\pi}=\frac{1}{6}(t_{\pi^0\pi^0\rightarrow\pi^0\pi^0}+2t_{\pi^0\pi^0\rightarrow\pi^+\pi^-}+2t_{\pi^+\pi^-\rightarrow\pi^0\pi^0}+4t_{\pi^+\pi^-\rightarrow\pi^+\pi^-})\ ,
\label{eq:isoampl}
\end{equation}
we can rewrite Eq. \eqref{eq:ampl} as
\begin{equation}
-it_{\sigma}=-i\ V^2\ \frac{3}{2}\ t_{\pi\pi\rightarrow\pi\pi}^{I=0}\ .
\label{eq:ampl2}
\end{equation}

Since the pions in the diagrams in Fig \ref{fig:sigmadiag} are off-shell, we need to use the off shell t-matrix obtained from the lowest order meson-meson Lagrangian \cite{oller}
\begin{equation}
t_{\pi\pi\rightarrow\pi\pi}^{I=0}=-\frac{1}{9f^2}\left(9s+\frac{15m_{\pi}^2}{2}-3\sum_ip_i^2\right)\ ,
\label{eq:offshellamp}
\end{equation}
with $s$ the Mandelstam variable and $m_{\pi}$ and $p_i$ the mass and momenta of the pions, respectively. As done in Ref. \cite{toki}, we can obtain the on-shell amplitude simply putting $p_i^2=m^2_{\pi}$ and this allow us to rewrite Eq. \eqref{eq:offshellamp} as 
\begin{equation}
t_{\pi\pi\rightarrow\pi\pi}^{I=0}=t_{\pi\pi\rightarrow\pi\pi}^{I=0,OS}+\frac{1}{3f^2}\sum_i(p_i^2-m_{\pi}^2)\ ,
\label{eq:offshellamp2}
\end{equation}
where
\begin{equation}
t_{\pi\pi\rightarrow\pi\pi}^{I=0,OS}=-\frac{1}{f^2}(s-\frac{m_{\pi}^2}{2})\ .
\label{eq:onshellamp}
\end{equation}

\begin{figure}[htpb]
\centering
\includegraphics[scale=0.5]{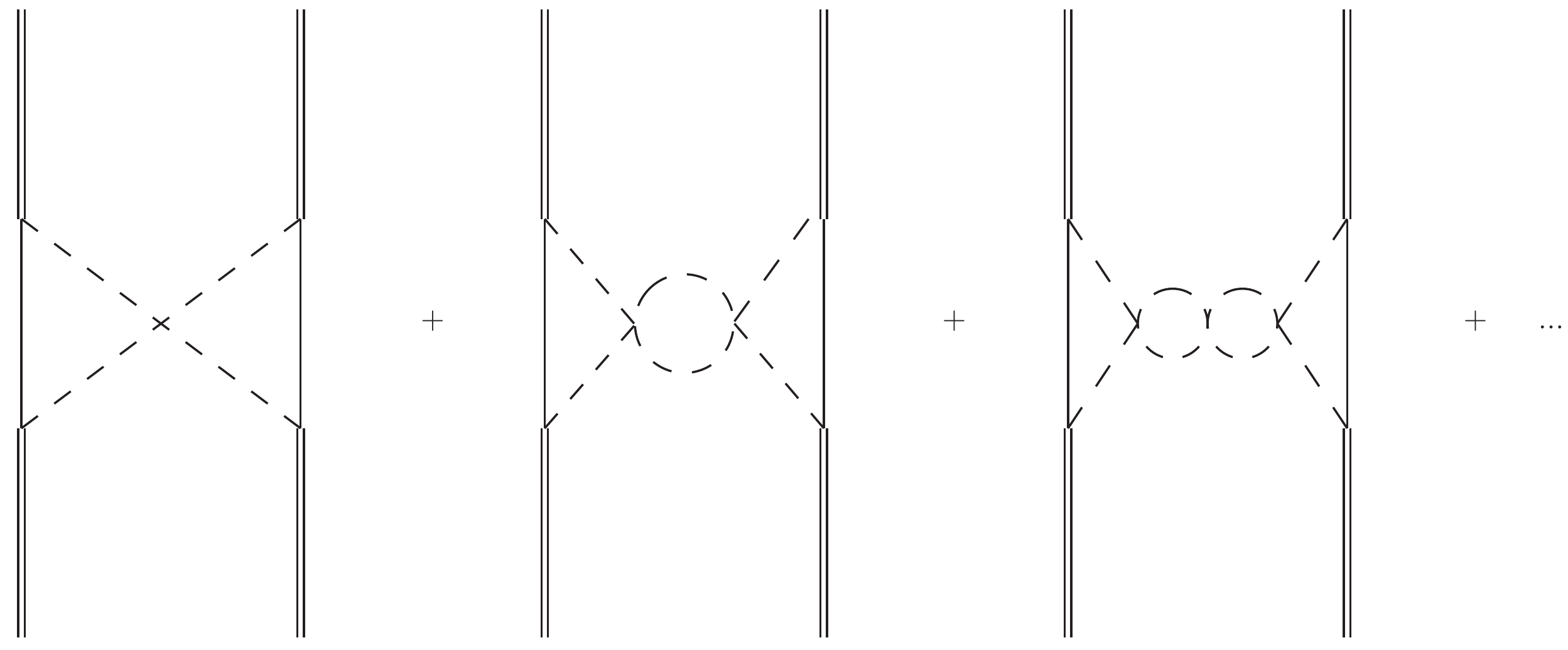}
\caption{$D\bar{D}^*$ interaction when the $\pi\pi$ scattering matrix is summed up to all orders in the unitary approach.}
\label{fig:sum}
\end{figure}

Following the approach of Ref. \cite{toki}, it can be shown that the off-shell part cancels exactly with other diagrams at  the same order in the chiral counting. Thus, at lowest order, we can write
\begin{equation}
t_{\sigma}=V^2\ \frac{3}{2}\ \frac{1}{f^2}(s-\frac{m_{\pi}^2}{2})\ .
\label{eq:ampl3}
\end{equation}
In order to apply the unitary Bethe-Salpeter approach to the scalar mesons amplitude, we need to sum  the set of diagrams in Fig. \ref{fig:sum}. This is easily done substituting the on-shell meson-meson amplitude of Eq. \eqref{eq:onshellamp} by \cite{oller}
\begin{equation}
t_{\pi\pi\rightarrow\pi\pi}^{I=0}=-\frac{1}{f^2}\ \frac{s-\frac{m_{\pi}^2}{2}}{1+\frac{1}{f^2}(s-\frac{m_{\pi}^2}{2})G(s)}\ ,
\label{eq:bs-ampl}
\end{equation}
where $G(s)$ is the two pions loop function, conveniently regularized \cite{toki},
\begin{equation}
G(s)=i\int \frac{d^4q}{(2\pi)^4}\,\frac{1}{q^2-m_{\pi}^2+i\epsilon}\,\frac{1}{(P-q)^2-m_{\pi}^2+i\epsilon}\ ,
\label{eq:loopf}
\end{equation}
with $P$ the total momentum of the two pion system and $P^2=s$.

\begin{figure}[htpb]
\centering
\includegraphics[scale=0.7]{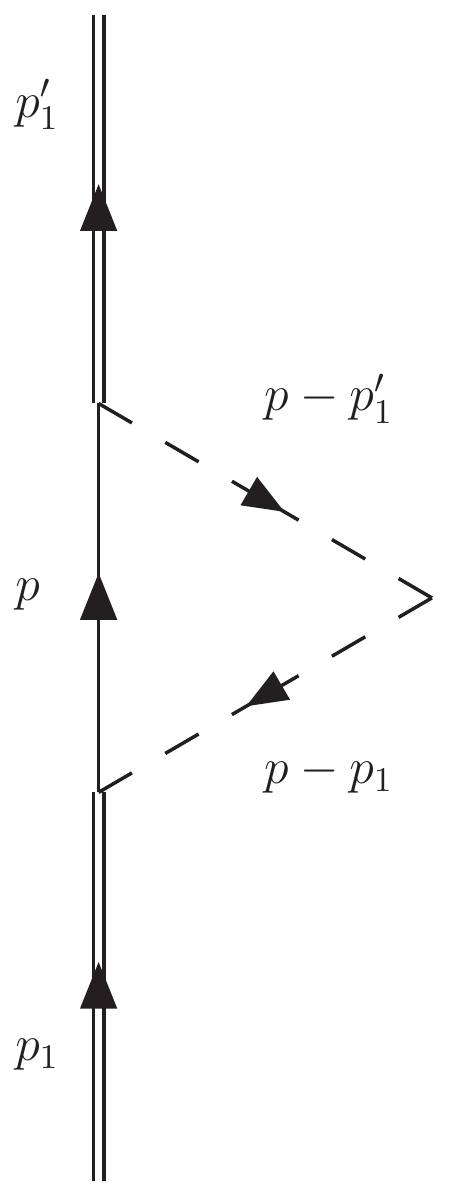}
\caption{Two pion exchange triangle vertex.}
\label{fig:V}
\end{figure}

We need, now, to evaluate the factor $V$ that appears in Eq. \eqref{eq:ampl3}, related, as already mentioned, to the triangular loop, which is shown in Fig. \ref{fig:V}. 
For simplicity, we use the Breit reference frame. This means that
\begin{equation}
\begin{split}
&p_1\equiv(p_1^0, \vec{q}/2)\ ,\\
&p_1'\equiv(p_1'^{\ 0}, -\vec{q}/2)\ ,\\
&p\equiv(p^0, \vec{p}\,)\ ,
\label{eq:breit}
\end{split}
\end{equation}
where $\vec{q}$ is the three-momentum transferred in the process. Since there is no energy exchange, $s=-\vec{q}^{\ 2}$. It is also useful to define the variable $q\equiv(0, \vec{q}\,)$.

Thus, by means of Eq. \eqref{eq:ampl_vert} and keeping in mind that we already factorized outside $V$ the coefficients $C$, we can write the expression of $V$ as
\begin{equation}
\begin{split}
V&=i\tilde{g}^2\int\frac{d^4p}{(2\pi)^4}\epsilon_{\mu}(2p-p_1)^{\mu}\epsilon'_{\nu}(2p-p_1')^{\nu}\frac {1}{p^2-m^2_D+i\epsilon}\\&\times\frac{1}{(p-p_1)^2-m^2_{\pi}+i\epsilon}\,\frac{1}{(p-p_1')^2-m^2_{\pi}+i\epsilon}\ ,
\label{eq:V}
\end{split}
\end{equation}
with $m_D$ the mass of the $D$ meson. Note that, as in section \ref{iterated}, we are using the coupling $\tilde{g}$ that accounts for the factor $m_{D^*}/m_{K^*}$ of the $HQSS$.

The integral in Eq. \eqref{eq:V} is logarithmically divergent. As in Ref. \cite{toki}, the regularization is accomplished by means of a cutoff in the space of intermediate states ($p_{max}=2$ GeV) and a form factor. In order to keep the integration in $p^0$ simple, we use the product of static form factors
\begin{equation}
F=F_1(\vec{p}+\frac{\vec{q}}{2})\, F_2(\vec{p}-\frac{\vec{q}}{2})=\frac{\Lambda^2}{\Lambda^2+(\vec{p}+\frac{\vec{q}}{2})^2}\,\frac{\Lambda^2}{\Lambda^2+(\vec{p}-\frac{\vec{q}}{2})^2}\ ,
\label{eq:ff}
\end{equation}
with $\Lambda=1$ GeV.

Since $\epsilon_{\mu}\,p_1^{\mu}=0$ and $\epsilon'_{\nu}\,p_1'^{\nu}=0$, Eq. \eqref{eq:V} can be rewritten as
\begin{equation}
\begin{split}
V&=4i\tilde{g}^2\int\frac{d^4p}{(2\pi)^4}\epsilon_{\mu}\,p^{\mu}\epsilon'_{\nu}\,p^{\nu}\frac {1}{p^2-m^2_D+i\epsilon}\,\frac{1}{(p-p_1)^2-m^2_{\pi}+i\epsilon}\,\frac{F}{(p-p_1')^2-m^2_{\pi}+i\epsilon}\ .
\label{eq:V2}
\end{split}
\end{equation}
The integral in Eq. \eqref{eq:V2} is symmetric with respect to $p_1$ and $p_1'$ and this allows us to derive the structure of the result of the integration, which will be of the type
\begin{equation}
V=\epsilon_{\mu}\epsilon'_{\nu}(ag^{\mu\nu}+b(p_1^{\mu}p_1^{\nu}+p_1'^{\mu}p_1'^{\nu})+c(p_1^{\mu}p_1'^{\nu}+p_1'^{\mu}p_1^{\nu}))\ .
\label{eq:struct}
\end{equation}

In the last expression, due to the Lorentz condition, only the terms $ag^{\mu\nu}$ and $cp_1'^{\mu}p_1^{\nu}$ survive but we need the entire structure to evaluate them. This is done taking the trace of Eq. \eqref{eq:V2} and multiplying the equation by $(p_{1\mu}p_{1\nu}+p_{1\mu}'p_{1\nu}')$ and $(p_{1\mu}p_{1\nu}'+p_{1\mu}'p_{1\nu})$, in order to obtain a system of three equations. Solving the system, we find the expressions of the three coefficients in Eq. \eqref{eq:struct} but, as we already said, we are only interested in
\begin{equation}
\begin{split}
&a=\frac{-Ym^2_{D^*}+Z(p_1p_1')+X(m^4_{D^*}-(p_1p_1')^2)}{2(m^4_{D^*}-(p_1p_1')^2)}\ ,\\
&c=\frac{-3Ym^2_{D^*}(p_1p_1')+X(m^4_{D^*}-(p_1p_1')^2)+Z(m^4_{D^*}+2(p_1p_1')^2)}{2(m^4_{D^*}-(p_1p_1')^2)^2}\ ,
\label{eq:ac}
\end{split}
\end{equation}
where
\begin{equation}
\begin{split}
&X=4\tilde{g}^2I_1+4\tilde{g}^2m_D^2I_2\ ,\\
&Y=8\tilde{g}^2p_1^{0\,2}I_1+8\tilde{g}^2I_3\ ,\\
&Z=8\tilde{g}^2p_1^{0\,2}I_1+8\tilde{g}^2I_4\ .
\label{eq:ac2}
\end{split}
\end{equation}

The four integrals in the equations above, $I_1$, $I_2$, $I_3$ and $I_4$, have the following expressions:
\begin{equation}
\begin{split}
&I_1=\int\frac{d^4p}{(2\pi)^4}\frac{1}{(p-p_1)^2-m_{\pi}^2+i\epsilon}\,\frac{1}{(p-p_1')^2-m_{\pi}^2+i\epsilon}\,F\ ,\\
&I_2=\int\frac{d^4p}{(2\pi)^4}\frac{1}{p^2-m_D^2+i\epsilon}\,\frac{1}{(p-p_1)^2-m_{\pi}^2+i\epsilon}\,\frac{1}{(p-p_1')^2-m_{\pi}^2+i\epsilon}\,F\ ,\\
&I_3=\int\frac{d^4p}{(2\pi)^4}\frac{(\vec{p}^{\,2}+m_D^2)p_1^{0\,2}+(\vec{p}\,\frac{\vec{q}}{2})^2}{p^2-m_D^2+i\epsilon}\,\frac{1}{(p-p_1)^2-m_{\pi}^2+i\epsilon}\,\frac{1}{(p-p_1')^2-m_{\pi}^2+i\epsilon}\,F\ ,\\
&I_4=\int\frac{d^4p}{(2\pi)^4}\frac{(\vec{p}^{\,2}+m_D^2)p_1^{0\,2}-(\vec{p}\,\frac{\vec{q}}{2})^2}{p^2-m_D^2+i\epsilon}\,\frac{1}{(p-p_1)^2-m_{\pi}^2+i\epsilon}\,\frac{1}{(p-p_1')^2-m_{\pi}^2+i\epsilon}\,F\ .
\label{eq:I}
\end{split}
\end{equation}

After performing the integration in $dp^0$, which can be done analytically using Cauchy's theorem, we obtain
\begin{equation}
\begin{split}
&I_1=\int\frac{d^3p}{(2\pi)^3}\,\frac{\omega_1+\omega_2}{2\omega_1\omega_2}\,\frac{1}{-\vec{q}^{\,2}-(\omega_1+\omega_2)^2}\,F\ ,\\
&I_2=\int\frac{d^3p}{(2\pi)^3}\,\frac{1}{2E_D}\,\frac{1}{2\omega_1}\,\frac{1}{\omega_2}\,\frac{1}{\omega_1+\omega_2}\,\frac{\omega_1+\omega_2+E_D-m_{D^*}}{E_D+\omega_1-m_{D^*}-i\epsilon}\,\frac{1}{E_D+\omega_2-m_{D^*}-i\epsilon}\,F\ ,\\
&I_3=\int\frac{d^3p}{(2\pi)^3}\,\frac{1}{2E_D}\,\frac{1}{2\omega_1}\,\frac{1}{\omega_2}\,\frac{1}{\omega_1+\omega_2}\,\frac{\omega_1+\omega_2+E_D-m_{D^*}}{E_D+\omega_1-m_{D^*}-i\epsilon}\,\frac{(\vec{p}^{\,2}+m_D^2)p_1^{0\,2}+(\vec{p}\,\frac{\vec{q}}{2})^2}{E_D+\omega_2-m_{D^*}-i\epsilon}\,F\ ,\\
&I_4=\int\frac{d^3p}{(2\pi)^3}\,\frac{1}{2E_D}\,\frac{1}{2\omega_1}\,\frac{1}{\omega_2}\,\frac{1}{\omega_1+\omega_2}\,\frac{\omega_1+\omega_2+E_D-m_{D^*}}{E_D+\omega_1-m_{D^*}-i\epsilon}\,\frac{(\vec{p}^{\,2}+m_D^2)p_1^{0\,2}-(\vec{p}\,\frac{\vec{q}}{2})^2}{E_D+\omega_2-m_{D^*}-i\epsilon}\,F\ ,
\label{eq:I2}
\end{split}
\end{equation}
where $\omega_1=\sqrt{(\vec{p}+\vec{q}/2)^2+m_{\pi}^2}$, $\omega_2=\sqrt{(\vec{p}-\vec{q}/2)^2+m_{\pi}^2}$ and $E_D=\sqrt{\vec{p}^{\,2}+m_D^2}$ are the energies of the two pions and of the $D$ meson involved in the loop, respectively, and $m_{D^*}$ is the mass of the $\bar{D}^*$ meson. Since the mass of the $D$ meson is so large, we have taken the positive energy component of the propagator $[(p^0-E_D)2E_D]^{-1}$, which simplifies the integration.

We can now go back to the $D^*\bar{D}^*$ potential in momentum space, whose final expression, according to Eqs. \eqref{eq:ampl2} and \eqref{eq:bs-ampl}, is given by
\begin{equation}
t_{\sigma}(\vec{q})=V^2\,\frac{3}{2}\,\frac{1}{f^2}\,\frac{\vec{q}^{\,2}+\frac{m_{\pi}^2}{2}}{1-G(-\vec{q}^{\,2})\,\frac{1}{f^2}(\vec{q}^{\,2}+\frac{m_{\pi}^2}{2})}\ ,
\label{eq:ampl-fin}
\end{equation}
with 
\begin{equation}
V=\epsilon_{\mu}\epsilon'_{\nu}(ag^{\mu\nu}+cp_1'^{\mu}p_1^{\nu})
\label{eq:V3}
\end{equation}
and $a$ and $c$ derived using Eqs. \eqref{eq:ac},  \eqref{eq:ac2} and  \eqref{eq:I2}. 

\begin{figure}[htpb]
\centering
\includegraphics[scale=0.4]{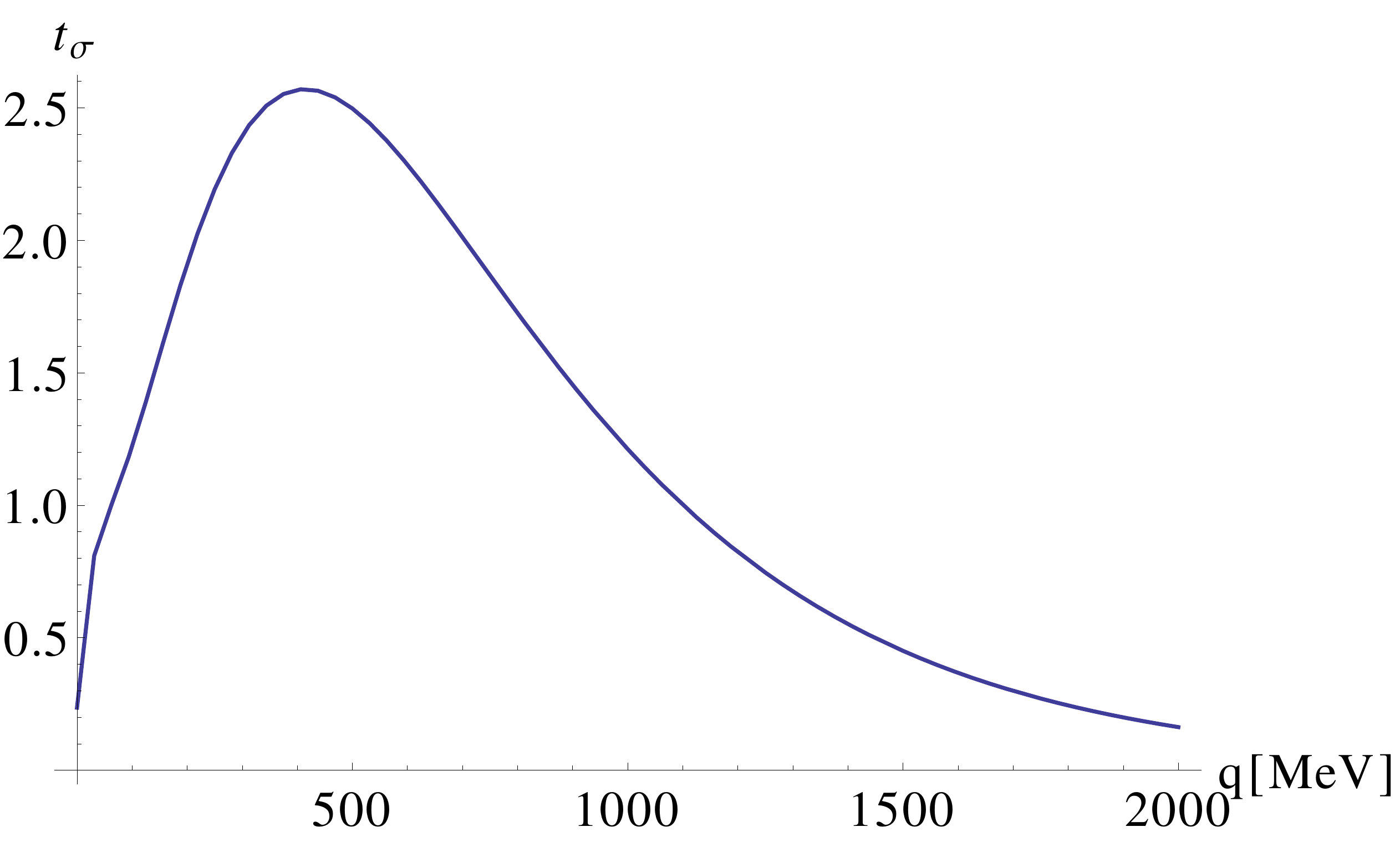}
\caption{Potential $t_{\sigma}$ as a function of the momentum transferred in the process.}
\label{fig:tsigma}
\end{figure}

Since we assume small initial momenta $\vec{p}_1$ and $\vec{p}_1\,'$ of the vectors compared to the vector mass, we can take $\epsilon^0\equiv 0$ and only the $a\epsilon\epsilon'$ combination remains. The other vertex will provide a similar structure. Hence, we have the combination
\begin{equation}
\epsilon^{(1)}_{i}\epsilon^{(2)}_{j}\epsilon^{(3)}_{i}\epsilon^{(4)}_{j}\ ,
\end{equation}
with $1+2\rightarrow 3+4$. 
\begin{equation}
\epsilon^{(1)}_{i}\epsilon^{(2)}_{j}\epsilon^{(3)}_{i}\epsilon^{(4)}_{j}\equiv \mathcal{P}^{(0)}+\mathcal{P}^{(1)}+\mathcal{P}^{(2)}\ .
\end{equation}
The strength of $t_{\sigma}(\vec{q}\,)$, removing $g^{\mu\nu}\epsilon_{\mu}\epsilon_{\nu}$, gives already the strength of the two pion exchange potential in $J=2$. The potential $t_{\sigma}$ as a function of the transferred momentum $\vec{q}$ is plotted in Fig. \ref{fig:tsigma}. Once again we see that this contribution is negligible compared to the vector exchange of Fig. \ref{fig:vecex}.

\subsection{Uncorrelated crossed two pion exchange}

Now we want to study the $D^*\bar{D}^*$ interaction when the pions exchanged are not interacting. In this case, only the diagrams $a)$ and $d)$ of Fig. \ref{fig:sigmadiag} contribute to the process. This means that the isospin factor, given by the different vertices involved, will be $\frac{5}{4}$, and we do  not have the $\pi\pi$ amplitude (see Eq. \eqref{eq:offshellamp}). Note that we take only the crossed diagrams. The iterated  one $\pi$ exchange (together with $\eta$ and $\eta^{\prime}$), which we saw was OZI suppressed, was already evaluated in section \ref{iterated} and we do not consider it.

\begin{figure}[htpb]
\centering
\includegraphics[scale=0.7]{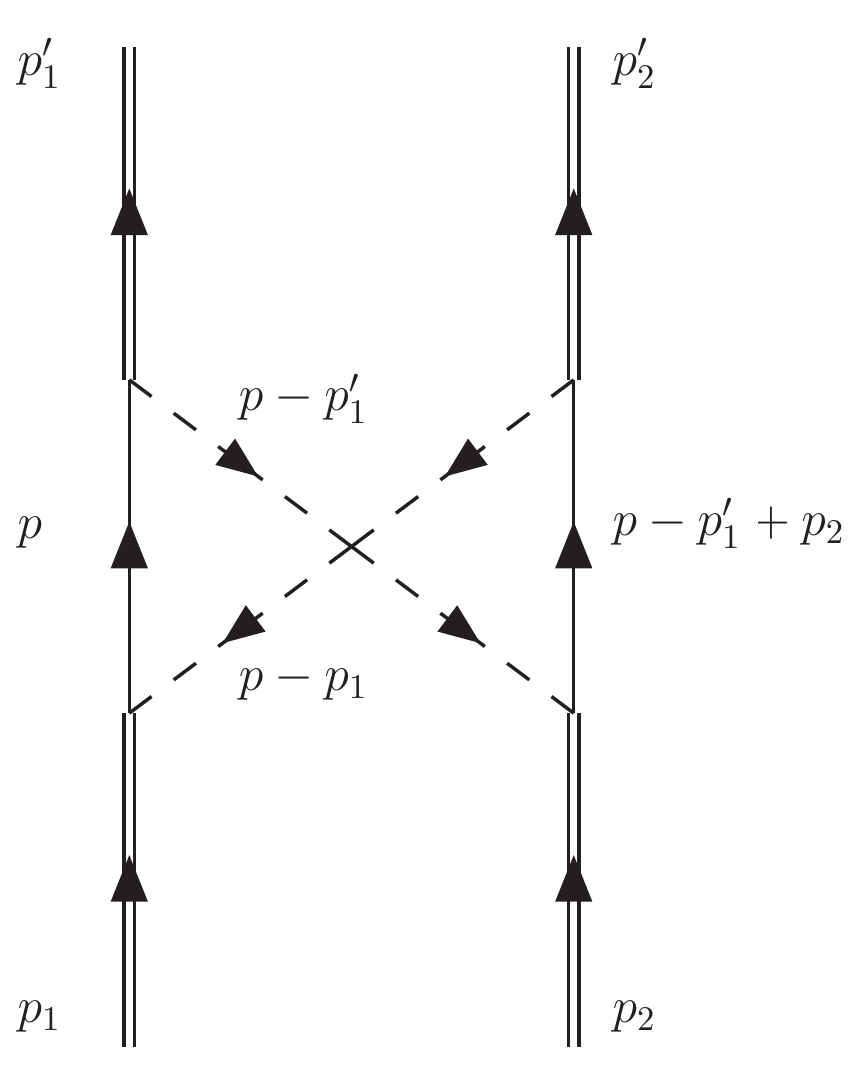}
\caption{Momenta assignment in the two pion exchange in $D\bar{D}^*\rightarrow D\bar{D}^*$.}
\label{fig:momenta}
\end{figure}

Recalling the expression of the vertices given in Eq. \eqref{eq:ampl_vert}, and choosing the momenta assignment as shown in Fig. \ref{fig:momenta}, we can directly write the amplitude of the process as
\begin{equation}
\begin{split}
t&=\frac{5}{4}i\tilde{g}^4\int\frac{d^4p}{(2\pi)^4}\,\epsilon_{\mu}(2p-p_1)^{\mu}\epsilon_{\nu}(2p-p_1')^{\nu}\epsilon_{\alpha}(2p-2p_1'+p_2)^{\alpha}\epsilon_{\beta}(2p-p_1'-p_1+p_2)^{\beta}\\&\times F^2\frac{1}{p^2-m^2_D+i\epsilon}\,\frac{1}{(p-p_1'+p_2)^2-m_D^2+i\epsilon}\,\frac{1}{(p-p_1)^2-m^2_{\pi}+i\epsilon}\\&\times\frac{1}{(p-p_1')^2-m^2_{\pi}+i\epsilon}\ .
\label{eq:ampl_rel}
\end{split}
\end{equation}

Since the momenta of the particles in the loop are small, we can use the non-relativistic approximation. The consequence is that only the spatial components of the polarization vectors survive and we can rewrite the amplitude of Eq. \eqref{eq:ampl_rel} as
\begin{equation}
\begin{split}
t&=\frac{5}{4}i\tilde{g}^4\int\frac{d^4p}{(2\pi)^4}\,\epsilon_{i}(2p-p_1)_{i}\epsilon_{j}(2p-p_1')_{j}\epsilon_{l}(2p-2p_1'+p_2)_{l}\epsilon_{m}(2p-p_1'-p_1+p_2)_{m}\\&\times F^2\frac{1}{p^2-m^2_D+i\epsilon}\,\frac{1}{(p-p_1'+p_2)^2-m_D^2+i\epsilon}\,\frac{1}{(p-p_1)^2+m^2_{\pi}+i\epsilon}\\&\times\frac{1}{(p-p_1')^2+m^2_{\pi}+i\epsilon}\ .
\label{eq:ampl_norel}
\end{split}
\end{equation}

We also assume that $4\vec{p}^{\,2}\gg\vec{q}^{\,2}/4$, such that the dominant term in Eq. \eqref{eq:ampl_norel}  is the one with the form $p_ip_jp_lp_m$. This means that  the amplitude in Eq. \eqref{eq:ampl_norel} will have the same structure as in the case of Section \ref{iterated}:
\begin{equation}
\frac{1}{15}\,(\delta_{ij}\delta_{lm}+\delta_{il}\delta_{jm}+\delta_{im}\delta_{jl})\ .
\label{eq:dominant}
\end{equation}

Thus, we can write
\begin{equation}
\begin{split}
t&=\frac{5}{4}i\tilde{g}^4\frac{1}{15}\int\frac{d^4p}{(2\pi)^4}(4\vec{p}^{\,2}-\frac{\vec{q}^{\,2}}{4})^2\,(\epsilon_{i}\epsilon_{l}\epsilon_{i}\epsilon_{l}+\epsilon_{i}\epsilon_{i}\epsilon_{l}\epsilon_{l}+\epsilon_{i}\epsilon_{l}\epsilon_{l}\epsilon_{i})\, F^2\frac{1}{p^2-m^2_D+i\epsilon}\\&\times\frac{1}{(p-p_1'+p_2)^2-m_D^2+i\epsilon}\,\frac{1}{(p-p_1)^2+m^2_{\pi}+i\epsilon}\,\frac{1}{(p-p_1')^2+m^2_{\pi}+i\epsilon}\ ,
\label{eq:ampl_norel2}
\end{split}
\end{equation}
that, performing the analytical integration in $dp^0$, becomes
\begin{equation}
\begin{split}
t&=\frac{5}{4}\tilde{g}^4\frac{1}{15}\int\frac{d^3p}{(2\pi)^3}(4\vec{p}^{\,2}-\frac{\vec{q}^{\,2}}{4})^2\,(\epsilon_{i}\epsilon_{l}\epsilon_{i}\epsilon_{l}+\epsilon_{i}\epsilon_{i}\epsilon_{l}\epsilon_{l}+\epsilon_{i}\epsilon_{l}\epsilon_{l}\epsilon_{i})\, F^2\,\frac{1}{\omega_1+\omega_2}\,\frac{1}{2\omega_1\omega_2}\\&\times\frac{1}{E_D^2} \left(1+\frac{E_D+\omega_1+\omega_2-p_1^0}{p_1^0-\omega_1-E_D+i\epsilon}+\frac{E_D+\omega_1+\omega_2-p_1^0}{p_1^0-\omega_2-E_D+i\epsilon}\right)\,\frac{1}{p_1^0-\omega_1-E_D+i\epsilon}\\&\times\frac{1}{p_1^0-\omega_2-E_D+i\epsilon}\ .
\label{eq:ampl_norel3}
\end{split}
\end{equation}

The combination of polarization vectors appearing in Eq. \eqref{eq:ampl_norel3} can be rewritten in terms of the spin projector operators  \cite{raquel} as
\begin{equation}
\epsilon_{i}\epsilon_{l}\epsilon_{i}\epsilon_{l}+\epsilon_{i}\epsilon_{i}\epsilon_{l}\epsilon_{l}+\epsilon_{i}\epsilon_{l}\epsilon_{l}\epsilon_{i}=5\mathcal{P}^{(0)}+2\mathcal{P}^{(2)}\ .
\label{eq:combination}
\end{equation}

\begin{figure}
  \centering
  \subfigure[]{\label{fig:pions0}\includegraphics[width=0.4965\textwidth]{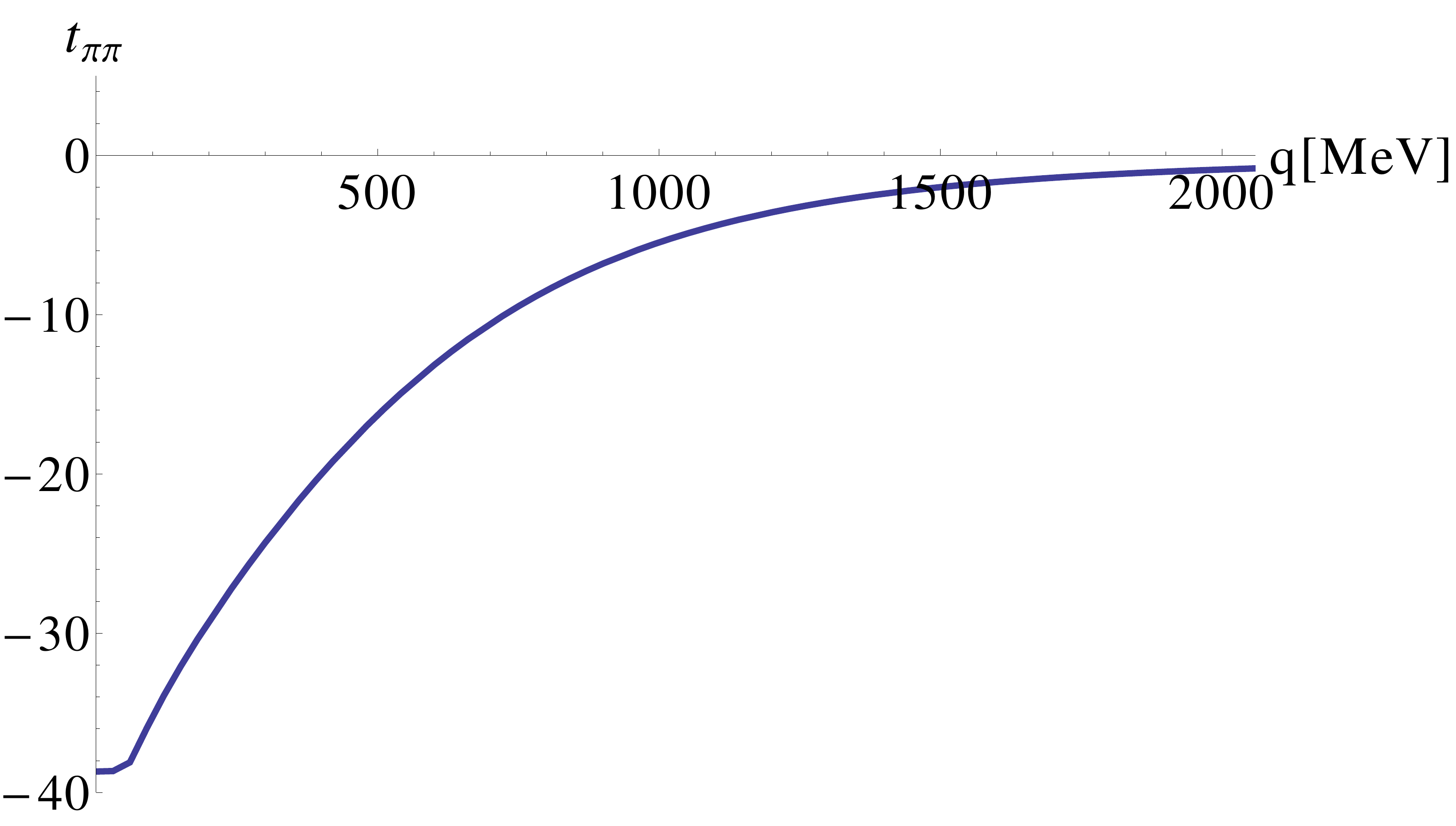}}                
   \subfigure[]{\label{fig:pions2}\includegraphics[width=0.4965\textwidth]{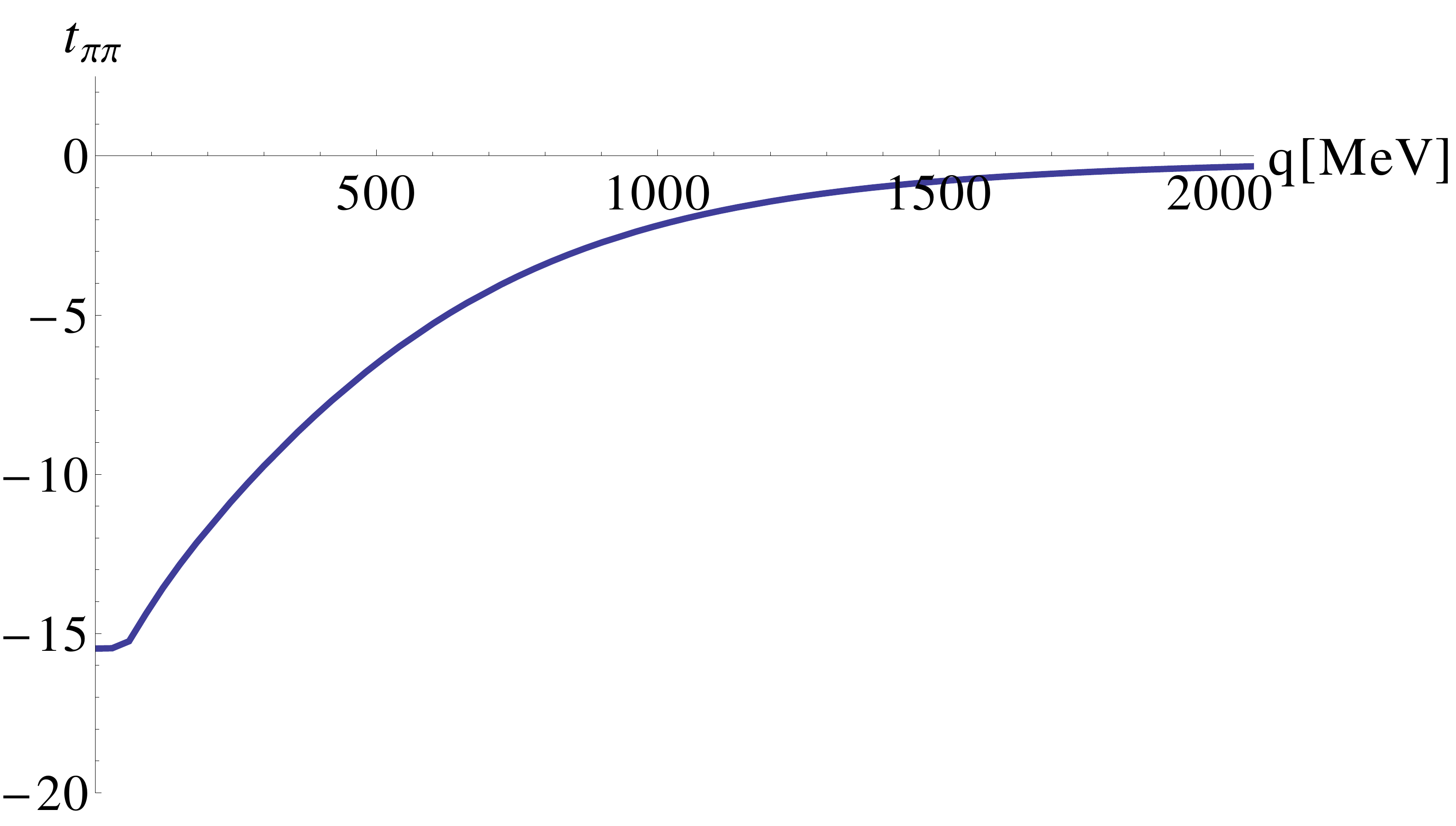}}
  \caption{Potential $t$ for non-interacting pion exchange in the case of $J=0$ $(a)$ and $J=2$ $(b)$.}
  \label{fig:pions}
\end{figure}

Thus, the final expression of the amplitude reads
\begin{equation}
\begin{split}
t&=\frac{5}{4}\tilde{g}^4\frac{A}{15}\int\frac{d^3p}{(2\pi)^3}(4\vec{p}^{\,2}-\frac{\vec{q}^{\,2}}{4})^2\, F^2\,\frac{1}{\omega_1+\omega_2}\,\frac{1}{2\omega_1\omega_2}\,\frac{1}{4E_D^2}\,\frac{1}{p_1^0-\omega_1-E_D+i\epsilon}\\&\times\frac{1}{p_1^0-\omega_2-E_D+i\epsilon} \left(1+\frac{E_D+\omega_1+\omega_2-p_1^0}{p_1^0-\omega_1-E_D+i\epsilon}+\frac{E_D+\omega_1+\omega_2-p_1^0}{p_1^0-\omega_2-E_D+i\epsilon}\right)\ ,
\label{eq:ampl_norel4}
\end{split}
\end{equation}
where $A=5$ for the $J=0$ case and $A=2$ for the $J=2$ case. The amplitude $t$ in the two cases is shown in Fig. \ref{fig:pions}. We can see that for $J=2$ the contribution is small compared with the one of vector exchange in Fig. \ref{fig:vecex}. Furthermore, we observe some cancellation between the repulsive one meson exchange of Fig. \ref{fig:tlightex} and the present contribution of Fig. \ref{fig:pions2}, and altogether we neglect all the terms coming from pseudoscalar exchange.

\section{Results}

In the former section we evaluated the contribution due to vector exchange and pseudoscalar exchange. We could see that the latter was small, of the order of $10\%$ or smaller than the other. In view of this, we should take this as an indicator that the strength of the vector exchange term can be changed by about $10\%$ when we evaluate uncertainties in our results.

\begin{figure}[htpb]
\centering
\includegraphics[scale=0.5]{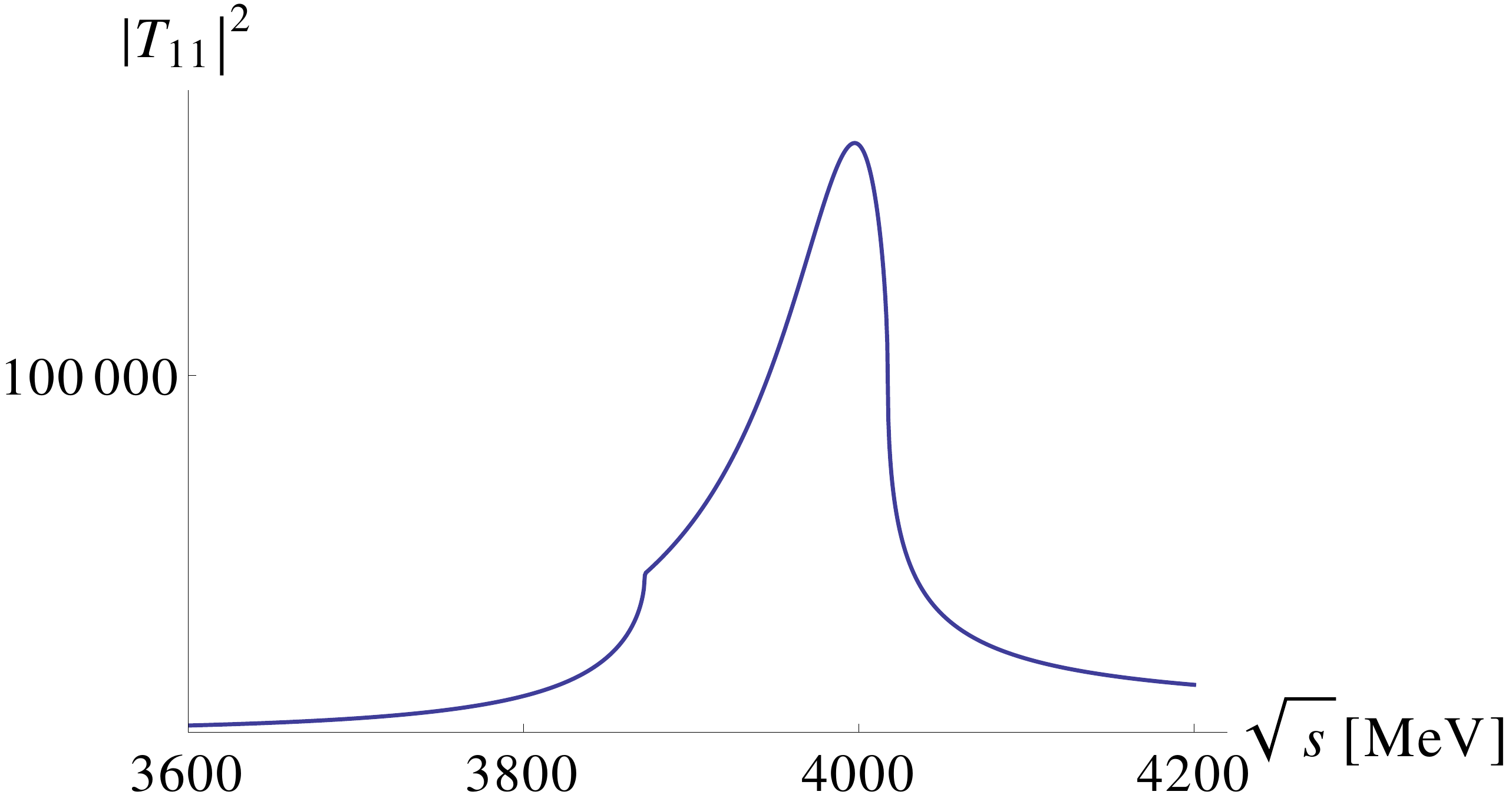}
\caption{$|T_{11}|^2$ as a function of $\sqrt{s}$.}
\label{fig:realpole}
\end{figure}

We want to study the $T$  matrix  for the two channels for values of $\sqrt{s}$ around $4000$ MeV. We study the shape of $|T|^2$. Fig. \ref{fig:realpole} shows $|T_{11}|^2$, where the subscript $11$ means that we are considering the transition from the channel $D^*\bar{D}^*$ to itself, as a function of the centre of mass energy. We use the dimensional regularization for the $G$ function (Eq. \eqref{eq:loopexdm}), choosing as the subtraction constants $\alpha_1=-2.3$ and $\alpha_2=-2.6$, while  $\mu=1000$ MeV. This is equivalent to using a cutoff $q_{max}=960$ MeV. With this choice of the parameters we obtain a clear peak around $\sqrt{s}=3998$ MeV, with a width $\Gamma\simeq 90$ MeV. This is about $19$ MeV below the $D^*\bar{D}^*$ threshold. The binding is smaller than found in \cite{raquel2} because we use $g$ for the coupling instead of $g_D$, which we justified from the findings of \cite{xiaoliang}.

This result is very interesting. Indeed, as mentioned in the Introduction,  in \cite{besexp} a peak was seen in the $(D^*\bar{D}^0)^{\pm}$ invariant mass spectrum close to the $(D^*\bar{D}^0)^{\pm}$ threshold, which was interpreted in \cite{besexp} as a signal of a $J=0$ resonance at $4025$ MeV.  However in \cite{alberdd} it was found that the spectrum could be equally reproduced assuming a $J=2$ resonance below threshold, with a mass around $3990$ MeV and a width of $160$ MeV. A fit with about $8$ MeV less binding and smaller width is also acceptable by looking at the different options discussed in \cite{alberdd}. Our choice of the parameters is  motivated to get a binding similar to that suggested in \cite{alberdd} but we discuss below our uncertainties. The finding of the present paper would give support to the interpretation of the results of \cite{besexp} as a consequence of an $I=1$ resonance coming from the $D^*\bar{D}^*$ interaction, with the option suggested in \cite{alberdd} of a bound $D^*\bar{D}^*$ state with relatively large width. 

\begin{figure}[htpb]
\centering
\includegraphics[scale=0.5]{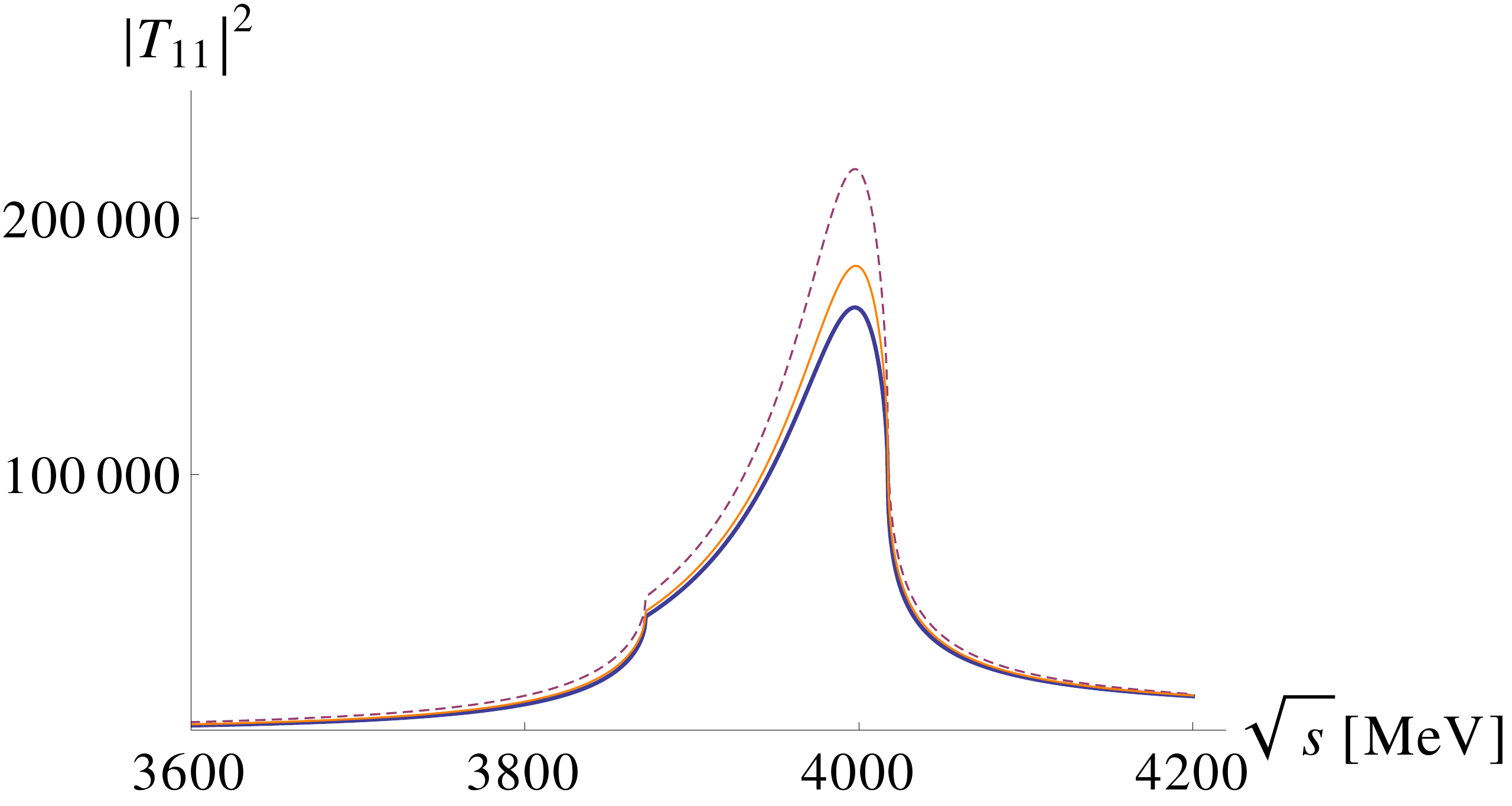}
\caption{$|T_{11}|^2$ as a function of $\sqrt{s}$ for the vector exchange potentials of Eq. \eqref{eq:DDDD} and \eqref{eq:DDRJ} (thick line), for an increase of $20 \%$ in the vector exchange potential (thin line) and for an increase of $50 \%$ (dashed line), for a peak at $3998$ MeV.}
\label{fig:error}
\end{figure}

We have also evaluated the uncertainties in the results due to the possible contribution of the two pion exchange in the interaction and the pseudoscalar one meson exchange ($\pi$, $\eta$, $\eta^{\prime}$). As we already mentioned in the beginning of this section, this contribution is small and attractive at small $\vec{q}$. In order to take it into account, we increase the magnitude of the vector exchange potential for $D^*\bar{D}^*\rightarrow D^*\bar{D}^*$ of Eq.  \eqref{eq:DDDD} and see how the position of the peak changes.  We find that, with an increase in the magnitude of the potential of $50\%$,  the energy of the peak decreases by about $5$ MeV.  Then we did the same thing, but adjusting the cutoff used in Eq. \eqref{eq:loopexdm}  in order to maintain fixed at $3998$ MeV the position of the peak. The results obtained are shown in Fig.  \ref{fig:error}. Increasing the magnitude of $t_{D^{*}\bar{D}^{*}\rightarrow D^{*}\bar{D}^{*}}$, the peak in $|T_{11}|^2$ is maintained in the same position using a lower cutoff. In  the case of an increase of $20\%$ (the thin, continuous line in Fig. \ref{fig:error}), we need a cutoff of  $q_{max}\simeq 940$ MeV, while in the case of $50 \%$ (the dashed line in Fig. \ref{fig:error}), $q_{max}\simeq 930$ MeV.  The shape of $|T_{11}|^2$ is slightly changed when going to higher magnitudes, giving a narrower peak and a higher strength.

\begin{figure}[htpb]
\centering
\includegraphics[scale=0.5]{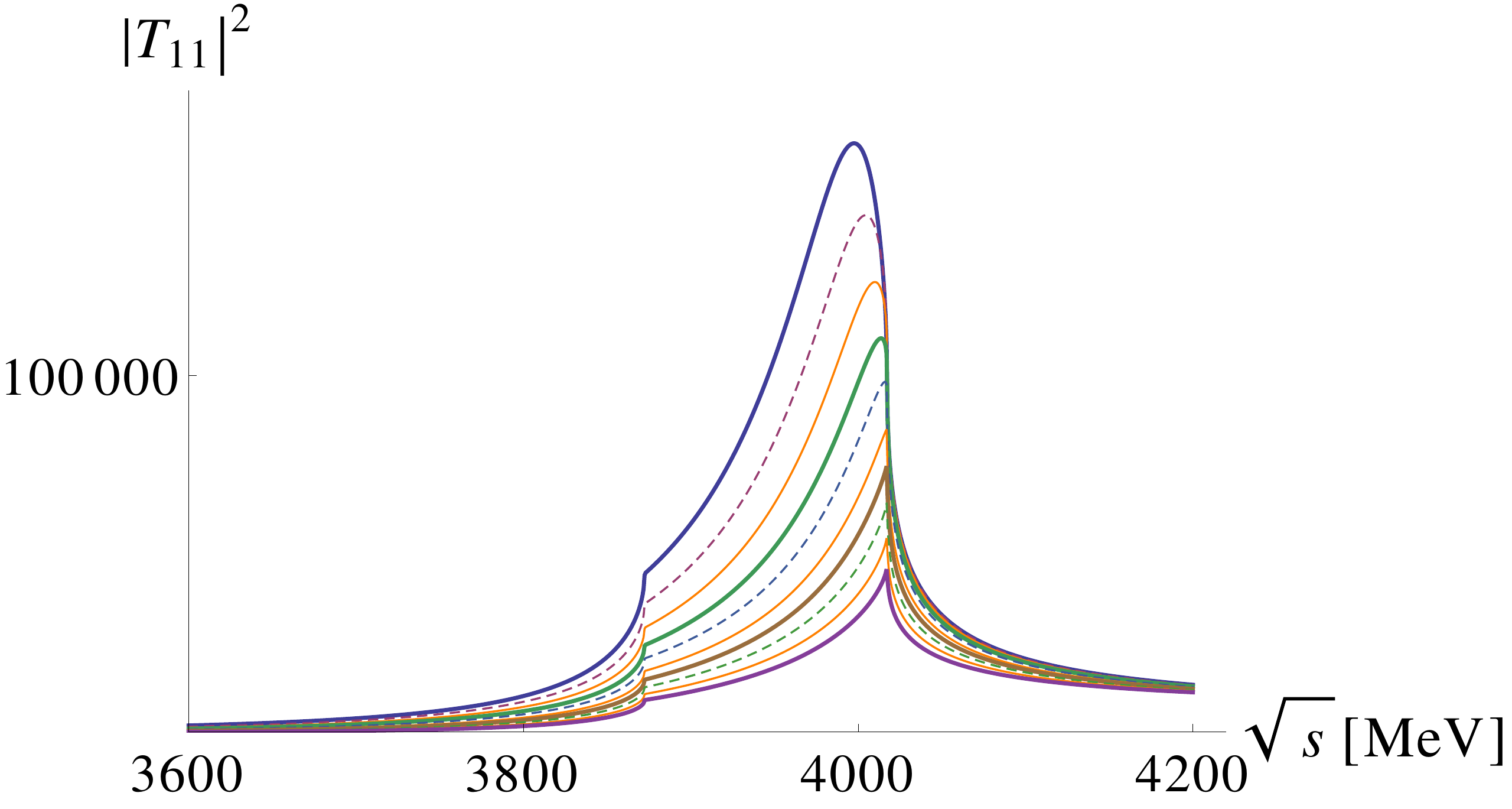}
\caption{$|T_{11}|^2$ as a function of $\sqrt{s}$, for different values of the cutoff $q_{max}$. From up down, $q_{max}=960,\,900,\,850,\,800,\,750,\,700,\,650,\,600,\,550,\,500$ MeV.}
\label{fig:cutoff}
\end{figure}

We have taken natural values for $\alpha_i$, or the cutoff, guided by the results of the analysis of \cite{alberdd}. Yet, it is interesting to see what happens if we reduce the cutoff.  In Fig. \ref{fig:cutoff} we show $|T_{11}|^2$ for different values of the cutoff.  We can see that as $q_{max}$ decreases, the peak of $|T_{11}|^2$ is moving closer to the threshold and its strength decreases. At  $q_{max}=700$ MeV we already have a clear cusp and, for lower values of $q_{max}$, the cusp remains but the strength of $|T_{11}|^2$ at the peak is very weak and we would no longer be able to produce an enhancement of the $D^* \bar D^*$ invariant mass distribution as seen in the experiment of \cite{besexp}. It is also interesting to see that even for values of $q_{max}\simeq 800$ MeV as in \cite{wuzou,wuraquel}, we still find a state bound by a few MeV. On the other hand, bigger values of $q_{max}$ would produce a too large binding that would contradict the results of the analysis of \cite{alberdd}. Hence, considering uncertainties in our model, we can say that we are obtaining a bound $D^*\bar{D}^*$ state or barely bound or even a virtual state (decaying to $J/\psi\rho$) within $3990-4000$ MeV, with a width of about $100$ MeV. Note that, even when the pole in the bound region gets close to threshold and disappears, it can get converted into a virtual state with a clearly visible cusp that can be translated into a peak close to threshold in an experimental analysis.

As already mentioned in Section \ref{vector}, in \cite{raquel2} also the $\rho\rho$, $\rho\omega$, $\rho\phi$ light vector channels were considered and  the $\rho\omega$ and $\rho\phi$ also give some contribution to the width. A slight increase in the value of $\Gamma\simeq 100$ MeV, would be in agreement with the analysis of \cite{alberdd} where $\Gamma=160$ MeV.

\section{Conclusions}
We have studied the interaction of $D^* \bar D^*$ in $I=1$ from the perspective of the local hidden gauge approach, extrapolating the model to account for the exchange of heavy vectors. This is necessary here once we prove that the exchange of light vectors is OZI forbidden in $I=1$. The interaction is then weaker than for $I=0$, where the exchange of light vectors is allowed, but still strong enough to weakly bind the system. We have also taken into account the coupled channel $J/\psi \rho$, which is open for decay and is responsible for a width of the state of the order of $100$ MeV. We also mention that the exchange of light $q \bar q$ is OZI forbidden, which implies that the sum of the exchange of light pseudoscalar mesons also vanishes if the masses of these mesons are taken degenerate. Because of that, we study the effect of two pion exchange, with and without interaction, but we find that this contribution is smaller than the exchange of heavy vectors. The study conducted here complements the one of \cite{alberdd} where the peak seen in the $D^* \bar D^*$ spectrum in the $e^+ e^- \to (D^* \bar D^*)^{\pm} \pi^{\pm}$ reaction, that led the experimental team to claim a $J^P=1^+$ $Z_c(4025)$, was reinterpreted as a possible $2^+$ bound state of  $D^* \bar D^*$ with $I=1$. Both the mass and width that we obtain are compatible with the results obtained in \cite{alberdd} from a fit to the experimental data, from where we would conclude that the state that we find in our approach can provide a natural explanation of the experimental results of \cite{besexp} and one could claim a resonance from this experiment but with a different energy, width ($M=3990-4000$ MeV, $\Gamma\simeq 100$ MeV) and quantum numbers ($I^G=1^-$, $J^{PC}=2^{++}$). 

\section*{Acknowledgments}
We would like to thank Fernando Navarra for useful comments.
This work is partly supported by the Spanish Ministerio de Economia y Competitividad and European FEDER funds under the contract number
FIS2011-28853-C02-01, and the Generalitat Valenciana in the program Prometeo, 2009/090. J. M. Dias acknowledges the Brazilian Funding Agency FAPESP for support. We acknowledge the support of the European Community-Research Infrastructure Integrating Activity Study of Strongly Interacting Matter (acronym HadronPhysics3, Grant Agreement
n. 283286) under the Seventh Framework Programme of EU.

\end{document}